\documentclass[onecolumn,showpacs]{revtex4}

\usepackage{amsmath}
\usepackage{graphicx}
\usepackage{dcolumn}
\usepackage{bm}

\input epsf

\begin{document}

\title{\Large Bouncing Universe in the Contexts of Generalized Cosmic Chaplygin Gas
and Variable Modified Chaplygin Gas}

\author{\bf~Tanwi~Bandyopadhyay$^1$\footnote{tanwi.bandyopadhyay@aiim.ac.in}
and~Ujjal~Debnath$^2$\footnote{ujjaldebnath@gmail.com,
ujjal@associates.iucaa.in}}

\affiliation{$^1$Adani Institute of
Infrastructure Engineering, Ahmedabad-382421, Gujarat, India.\\
$^2$Department of Mathematics, Indian Institute of Engineering
Science and Technology, Shibpur, Howrah-711103, India.}

\begin{abstract}
In this work, we have considered the Friedmann-Robertson-Walker
(FRW) model of the universe where bounce occurs and the universe
is filled with Generalized Cosmic Chaplygin Gas (GCCG) or Variable
Modified Chaplygin Gas (VMCG). We have studied the stability
analysis through dynamical system for both models and found the
critical points in flat, open and closed universe. In presence of
scalar field, the dynamical behavior of scale factor and Hubble
parameter is described in both models. Finally, we have analyzed
the energy conditions for both the models in bouncing universe.\\


\end{abstract}


\maketitle

\section{\normalsize\bf{Introduction}}

The initial singularity in the cosmological models is the big
problem in the background of General Relativity (GR). The
inflation theories suffer from the initial singularity problem
\cite{Bor} at the origin $t=0$. The inflationary cosmology for
early-time is consistent but many theoretical inconsistencies of
the big bang cosmology had occurred \cite{Lin,Lyth,Gor}.
Inflationary paradigm may resolve several kinds of problems in
standard big bang cosmology in the early universe
\cite{Mukha,Sato} but the initial singularity cannot be avoided
\cite{Brand}. Since the inflation is described by the dynamics of
the scalar fields coupled with the Einstein's gravity
\cite{Brand1}, so a new gravitational theory may be required to
describe the beginning of the universe to avoid the initial
singularity. Many researchers have attempted to resolve this
singularity problem through the generalized/modified general
relativity theory \cite{Nov} say for example, Superstring theory,
loop quantum gravity etc. The initial singularity can be avoided
in frames of non-singular bouncing cosmological models \cite{Muk}.
The key feature of such models is modification of standard
Einstein-Hilbert action. The oscillating universe \cite{Tol} is an
alternative to standard big bang cosmology \cite{Mis,Peeb,Wald} to
avoid the big bang singularity by replacing it with a cyclical
evolution. A bouncing universe has an initial narrow state by a
minimal radius and then develops to an expanding phase \cite{Cai},
which means that the universe was arriving to the Big Bang era
after the bouncing and so the equation of state parameter should
cross from $\omega<-1$ to $\omega>-1$. Several studies in bouncing
cosmology could be found in literatures
\cite{Cai0,Nojiri0,Odin0,Sad,Ant,Xu0,Geg}. On the other hand, the
bouncing cosmology has also been studied in the framework of some
modified theories of gravity eg. $f(R)$ gravity
\cite{Ghana,Fard,Carl,Paul0,Barra,Odin,Odin1}, $f(T)$ gravity
\cite{Asta,Cai00}, Gauss-Bonnet gravity \cite{Bam,Oik},
Brans-Dicke theory \cite{Singh}, braneworld gravity
\cite{Stein0,Stein1,Piao0,Piao1} and loop quantum cosmology
\cite{Boj,Sin,Haro}.\\

Several works on extended Chaplygin gas have been done in
references \cite{P1,P2,P3,P4,P5,P6,P7,P8}. The variable
generalized and modified Chaplygin gas models have been discussed
in \cite{Q1,Q2,Q3}. Moreover there are some other models like
viscous Chaplygin gas, cosmic Chaplygin gas discussed in
\cite{R1,R2,R3,R4,R5,R6,R7,R8,R9,R10,R11,R12,R13,R14,R15,R16}.
Recently Salehi \cite{Salehi} has studied the bouncing universe in
presence of extended Chaplygin gas in the framework of
Friedmann-Robertson-Walker (FRW) model. Motivated by the work, we
have investigated the bouncing universe model in generalized
cosmic Chaplygin gas (GCCG) and variable modified Chaplygin gas
(VMCG). We have organized the paper in the following way. In
section II, we have considered the bouncing framework of FRW model
with GCCG or VMCG. In section III, we have studied the dynamical
system analysis in both cases and shown the phase space analysis.
In Sec. IV, we have discussed the possibility of an oscillating
universe in the presence of a scalar field. In section V, we have
examined the null energy conditions (NEC). Finally, we have drawn
the conclusion in section VI.

\section{\normalsize\bf{Main Equations in FRW Universe and Bounce Conditions}}

We assume the non-flat FRW model of the universe which is given by

\begin{equation}
ds^{2}=-dt^{2}+a^{2}(t)\left[\frac{dr^{2}}{1-kr^{2}}+r^{2}(d\theta^{2}+\sin^{2}\theta
d\phi^{2})\right]
\end{equation}
where $a(t)$ is the scale factor and  $k ~(=0, ~\pm 1)$  is the
curvature index of the universe described as flat, closed and open respectively.\\

If the FRW universe is undergoing a `bounce' then it attains a
minimum and at this minimum, the strong energy condition (SEC) of
classical gravity must be violated (which is necessary but not the
sufficient condition) \cite{Singh}. In a bouncing universe, the
universe first undergoes a collapse to attains a minimum and then
subsequently expands. For the bounce in FRW model, during
contraction phase of the universe, $a(t)$ decreases i.e.,
($\dot{a}(t) < 0$) and then in the expanding phase of the
universe, $a(t)$ increases i.e., ($\dot{a}(t) > 0$). At the bounce
point, $t = t_{b}$, the minimal necessary condition is
$\dot{a}(t)=0$ and $\ddot{a}(t) > 0$ for $t\in (t_{b}-\epsilon,
t_{b})\cup (t_{b},t_{b}+\epsilon)$ for small $\epsilon>0$. For a
non-singular bounce $a(t_{b})\ne 0$. The conditions may not be
sufficient for a non-singular bounce.\\

In non-flat FRW model, the Einstein's field equations are given by
(assuming $8\pi G=1$)
\begin{equation}
3\left(H^2+\frac{k}{a^2}\right)=\rho,
\end{equation}
\begin{equation}
\left(2\dot{H}+3H^2+\frac{k}{a^2}\right)=-p
\end{equation}
where $H=\frac{\dot{a}}{a}$ is the Hubble parameter. At the bounce
time $t_{b}, H = 0$ and $\dot{H} > 0$ in a small time interval
near the bounce time. For $\dot{H} > 0$, we have $\rho+3p<0$, so
SEC must be violated. The energy conservation equation is given by

\begin{equation}
\dot{\rho}+3H(\rho+p)=0
\end{equation}

Now let us assume that the universe is filled with generalized
cosmic Chaplygin gas (GCCG) or variable modified Chaplygin gas
(VMCG). So here $\rho$ and $p$ represent respectively the energy
density and pressure of GCCG or VMCG. The equations of state for
the two models GCCG \cite{Gonz,Chak} and VMCG \cite{Deb} are
respectively

\begin{equation}\label{EoS GCCG}
p_{_{GCCG}}=-\rho_{_{GCCG}}^{-\alpha}\left[C+(\rho_{_{GCCG}}^{1+\alpha}-C)^{-\omega}\right]
\end{equation}
and
\begin{equation}\label{EoS VMCG}
p_{_{VMCG}}=A\rho_{_{VMCG}}-\frac{B(a)}{\rho_{_{VMCG}}^{\alpha}}
\end{equation}

where $C=\frac{\tilde{A}}{1+\omega}-1$ with $\tilde{A}$ a constant
which can take on both positive and negative values and $-l<w<0$,
$l$ being a positive definite constant which can take on values
larger than unity, $A>0$ and $0\leq\alpha\leq1$. Also $B(a)=B_0
a^{-n}$ \cite{Deb}, with $B_0>0$ and $n$ is a positive constant.
It should be noted that VMCG starts from radiation era but GCCG
starts from dust era. Thus the solutions for the two different
models GCCG and VMCG become

\begin{equation}\label{rho gccg}
\rho_{_{GCCG}}=\left[C+\left(1+\frac{D}{a^{3(1+\alpha)(1+\omega)}}\right)
^{\frac{1}{1+\omega}}\right]^{\frac{1}{1+\alpha}}
\end{equation}
and
\begin{equation}\label{rho vmcg}
\rho_{_{VMCG}}=\left[\frac{3(1+\alpha)B_0}{a^n[3(1+\alpha)(1+A)-n]}
+\frac{E}{a^{3(1+A)(1+\alpha)}}\right]^{\frac{1}{1+\alpha}}
\end{equation}

where $D>0$ and $E>0$ are integration constants with $3(1+\alpha)(1+A)>n$.\\

\section{\normalsize\bf{Dynamical System and Stability Analysis}}

A theoretically viable model under investigation can be conceived
easily by the study of its dynamical system. It means that the
analytical solutions for the model needs to be stable (or
asymptotically stable under small perturbations). This can be
visualized from the phase space analysis and the behaviour of the
system can be well-understood. Many works exist in the literature
where extensive studies were performed to different dark energy
models by forming their dynamical systems and further analysing
their stability criteria. In \cite{Lopez}, a phantom scalar field
was studied with a non-coupled perfect fluid having a constant
equation of state. Inhomogeneous dust with a positive cosmological
constant given by the Lema$\hat{i}$tre-Tolman-Bondi (LTB) model
was investigated in \cite{Sussman1} and the critical points were
examined in a unique way. Further recent studies in the
inhomogeneous sector was conducetd with interactive mixture of
dark fluids \cite{Sussman2}. Interactive holographic dark energy
models were explored in the light of stability analysis in
Einstein's gravity \cite{Biswas} and other theories of gravity
\cite{Chakraborty}. A case study of tachyonic scalar field dark
energy, non-minimally coupled to the Barotropic fluid as the
matter of the universe, was carried out in a novel way by best
fitting the stability parameters to the observational data
\cite{Farajollahi}. Very recently, a bouncing Bianchi-IX model was
proposed in Ho$\tilde{r}$ava-Lifshitz gravity \cite{Maier} with a
positive cosmological constant along with non-interacting dust and
radiation as the dark energy and matter component.\\

In the present work, we propose to perform a systematic stability
analysis for both GCCG and VMCG models and conclude about their
theoretical viability. In order to do so, let us introduce the
following variables:

\begin{equation}
\chi=H,~~~~\zeta=a,~~~~\eta^2=\rho
\end{equation}

Then from the field equations with proper substitution of the
expressions of energy densities for both cases, we get the final
expressions of the autonomous system as

\begin{equation}\label{chai gccg}
\dot{\chi}|_{_{GCCG}}=\frac{k}{\zeta^2}+\frac{\left[C+\left\{\left(3\chi^2+\frac{3k}{\zeta^2}\right)^{(1+\alpha)}
-C\right\}^{-\omega}\right]
-\left(3\chi^2+\frac{3k}{\zeta^2}\right)^{(2+\alpha)}}{2\left(3\chi^2+\frac{3k}{\zeta^2}\right)^{(1+\alpha)}}
\end{equation}

\begin{equation}\label{chai vmcg}
\dot{\chi}|_{_{VMCG}}=-\frac{k(1+3A)}{2\zeta^2}-\frac{3(1+A)}{2}\chi^2
+\frac{B_0}{2~3^{\alpha}\zeta^{n}\left(\chi^2+\frac{k}{\zeta^2}\right)^{\alpha}}
\end{equation}
and
\begin{equation}\label{zeta eqn}
\dot{\zeta}=\zeta\chi
\end{equation}

Equation \eqref{zeta eqn} is applicable for both the models.
Solving them, the critical point $(\chi_c,\zeta_c)$ can be
obtained as

\begin{equation}
\chi_c=0
\end{equation}

and $\zeta_c$ is the root of the equations

\begin{equation}\label{root eqn GCCG}
3\zeta_c^{2(2+\alpha)}\left[C+\left\{(3k)^{1+\alpha}\zeta_c^{-2(1+\alpha)}-C\right\}^{-\omega}\right]=(3k)^{2+\alpha}
\end{equation}

and

\begin{equation}\label{root eqn VMCG}
B_0\zeta_c^{(2\alpha-n)+2}=(3k)^{\alpha+1}k(1+3A)
\end{equation}

for the GCCG and the VMCG models respectively.\\

It can be seen that the critical points depend on a number of
parameters for both the models. In case of GCCG, both $\alpha$ and
$\omega$ are crucial for the determination of critical points
whereas in the case of VMCG, both $\alpha$ and $n$ play major
roles. Here we shall study the dynamical system of equations
\eqref{chai gccg}, \eqref{chai vmcg} and \eqref{zeta eqn} to
obtain the critical points of the system. In this context, the
Jacobian Matrices for the two models are (for $\chi_c=0$)

\begin{equation}
J_1=\left(
\begin{array}{cc}
   0 &  \frac{U}{3k\zeta_c^{3}}\\
   \zeta_c & 0 \
 \end{array}
 \right)
\end{equation}

and

\begin{equation}
J_2=\left(
\begin{array}{cc}
   0 ~~~~~~~&
   \frac{(1+3A)k}{\zeta_c^{3}}-\frac{B_o(n-2\alpha)}{2{(3k)}^{\alpha}\zeta_c^{(n-2\alpha)+1}}\\\\
   \zeta_c ~~~~~~~~& 0 \
 \end{array}
 \right)
\end{equation}

where $U$ is given by

\begin{equation}
U=3k^2+\zeta_c^{2}(1+\alpha)P^{-(\omega+1)}\left[(3k)^{-\alpha}\zeta_c^{2(\alpha+1)}
P\left(1+CP^{\omega}\right)+3k\omega\right]
\end{equation}

with

\begin{equation}
P=-C+(3k)^{1+\alpha}{\zeta_c}^{-2(1+\alpha)}
\end{equation}

Hence the eigenvalues for the two models are given by

\begin{equation}
\lambda|_{_{GCCG}}=\pm\frac{1}{\zeta_c}\sqrt{\frac{U}{3k}}
\end{equation}

and

\begin{equation}
\lambda|_{_{VMCG}}=\pm\frac{\sqrt{k}}{\zeta_c}\sqrt{(1+3A)
+\frac{B_0(2\alpha-n){\zeta_c}^{(2\alpha-n)+2}}{2~3^{\alpha}k^{\alpha+1}}}
\end{equation}

For the GCCG model, we can not state that $U$ is positive definite
as it's value depends on many parameters viz. $\alpha$, $\omega$,
$k$ and $A$. Therefore the only pair of complex roots with the
zero real part do not come by assuming $k=-1$ only. In the VMCG
model, for the case of $\alpha=1$, if $n\leq2$, i.e., either $n=1$
or $n=2$, then for $k=-1$, we shall obtain two stable critical
points in the phase space of the system.\\

In the following the critical points are analyzed for different
values of the parameters involved. The critical points can be
obtained analytically from equations \eqref{root eqn GCCG} and
\eqref{root eqn VMCG} respectively for the two models. For GCCG
model, analytical expressions are too complicated to express here,
so we provide the analysis by studying the phase space diagrams
for both the models. Note that, in this model, the case of
$\omega=-1$ should be avoided as the term $C$ diverges in
this case.\\

\subsection{\normalsize\bf{Case 1: $k=-1,~~\alpha=1/2$}}

In GCCG model, we do not get any stable critical points assuming
$\omega=-0.5$ and $C=1$. Similar solution appears in VMCG model as
well when we choose $A=0, n=2$ and $B_0=0.5$. For GCCG and VMCG
model, the graphs of $a$ vs
$H$ are shown in Figs. 1 and 2 and $a$, $H$, $\rho$ vs $t$ are shown in Figs. 3-8 respectively.\\

\begin{figure}
\includegraphics[height=2.0in]{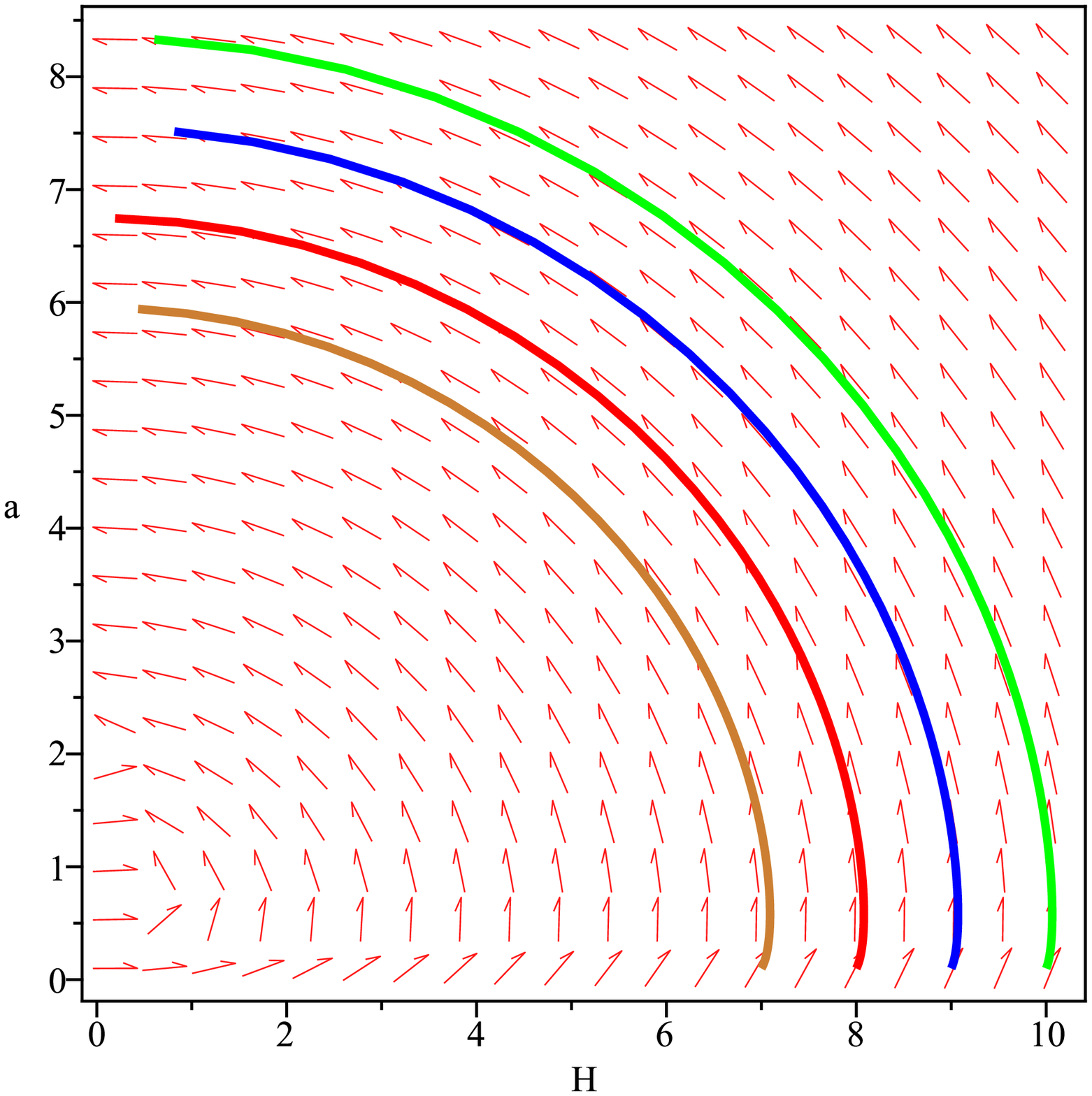}~~~~
\includegraphics[height=2.0in]{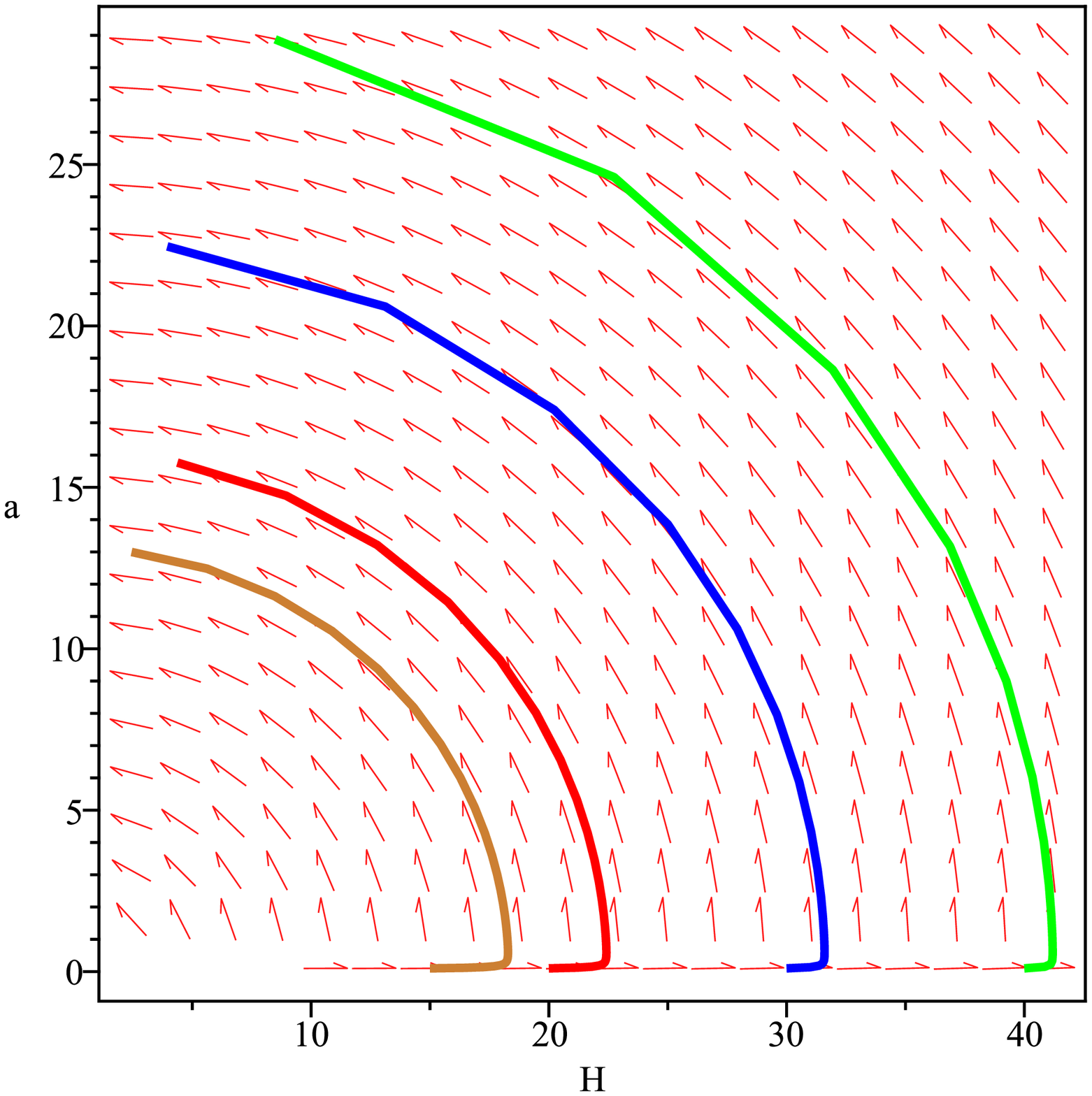}\\

\vspace{2mm} ~~~~~~~~~~~~Fig.1~~~~~~~~~~~~~~~~~~~~~~~~~~~~~~~~~~~~~~~~~~~Fig.2~~~~\\
\vspace{4mm}

\includegraphics[height=2.0in]{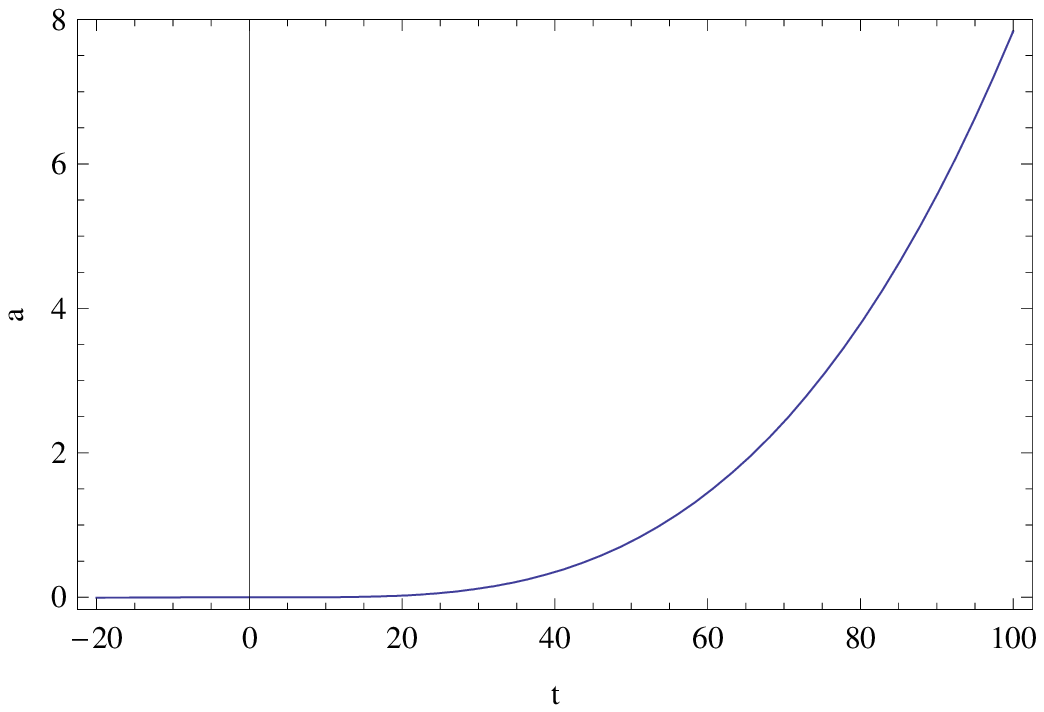}~~~~
\includegraphics[height=2.0in]{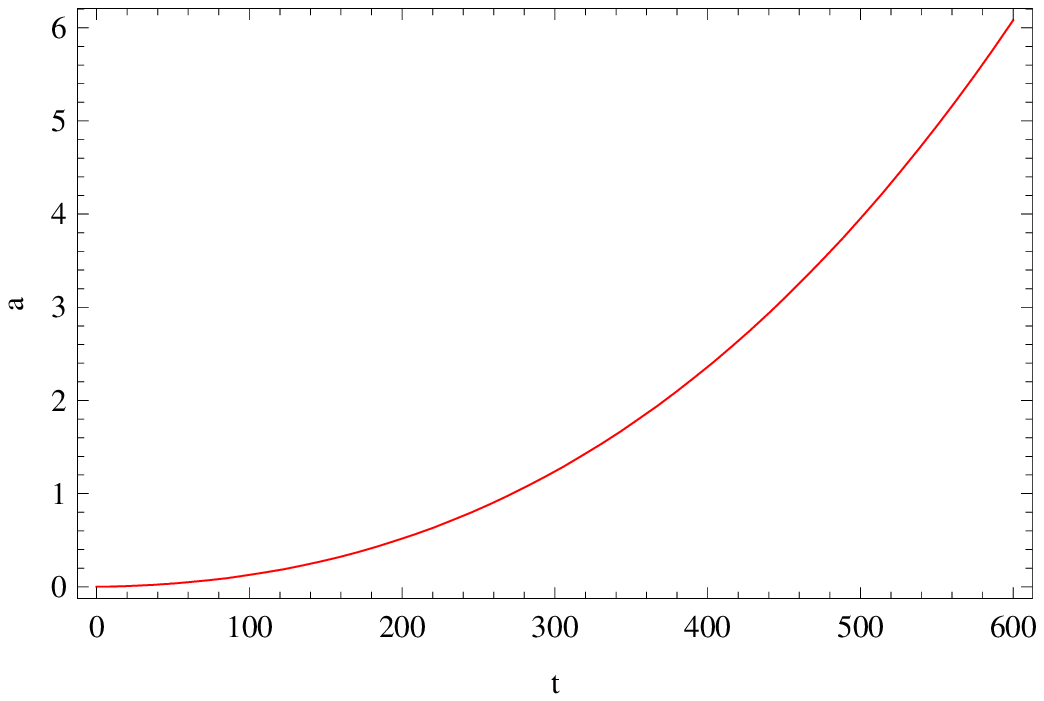}\\

\vspace{2mm} ~~~~~~~~~~~~Fig.3~~~~~~~~~~~~~~~~~~~~~~~~~~~~~~~~~~~~~~~~~~~Fig.4~~~~\\
\vspace{4mm}

\includegraphics[height=2.0in]{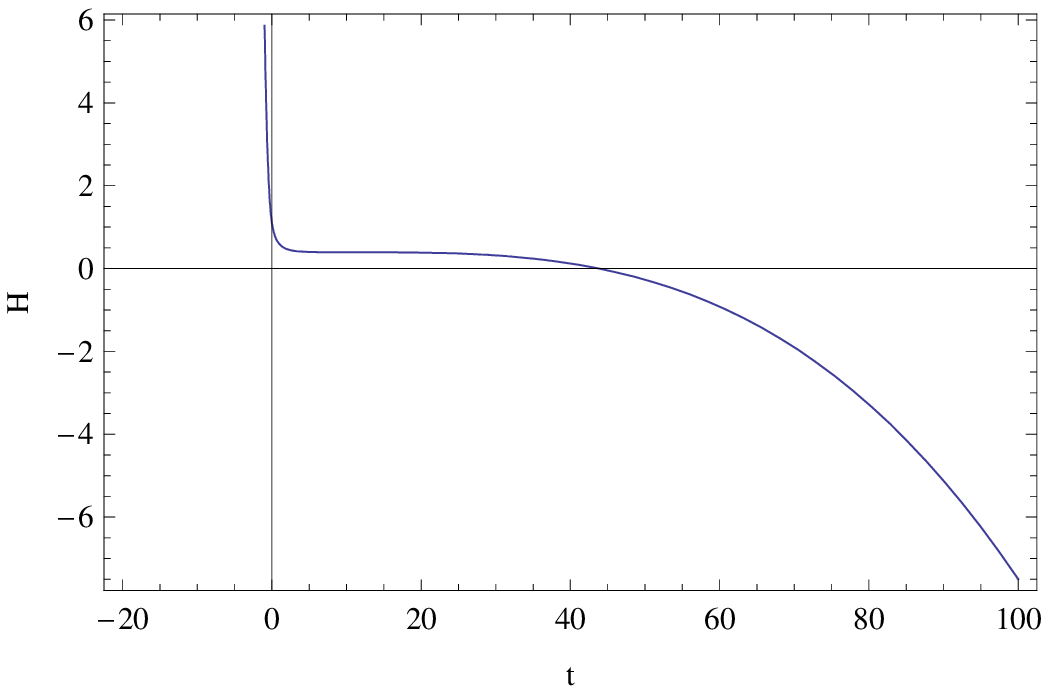}~~~~
\includegraphics[height=2.0in]{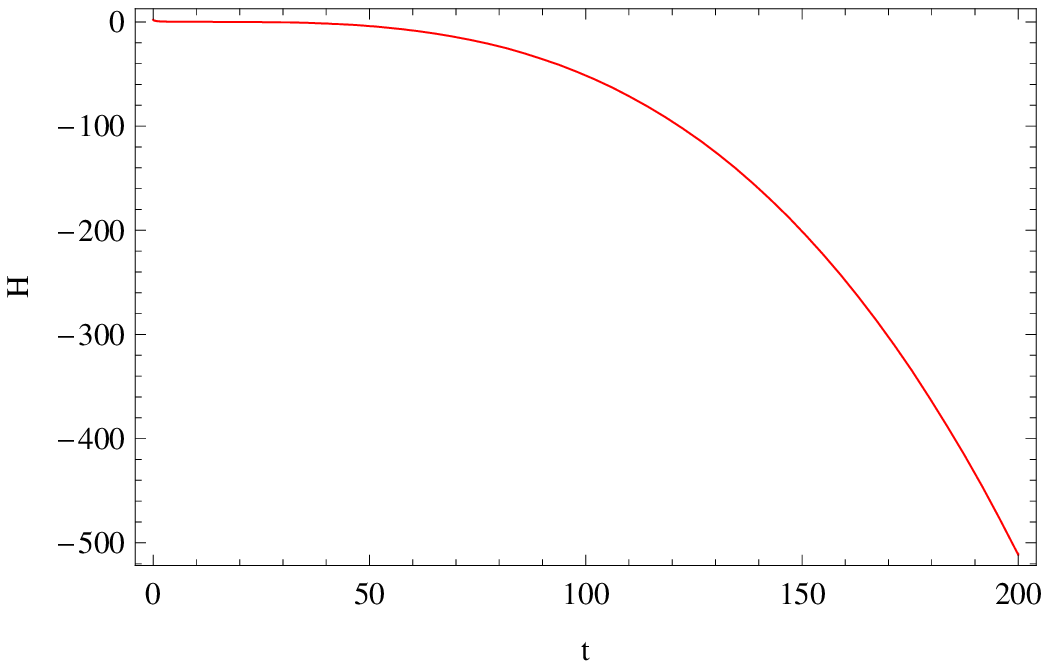}\\

\vspace{2mm} ~~~~~~~~~~~~Fig.5~~~~~~~~~~~~~~~~~~~~~~~~~~~~~~~~~~~~~~~~~~~Fig.6~~~~\\
\vspace{4mm}

\includegraphics[height=2.0in]{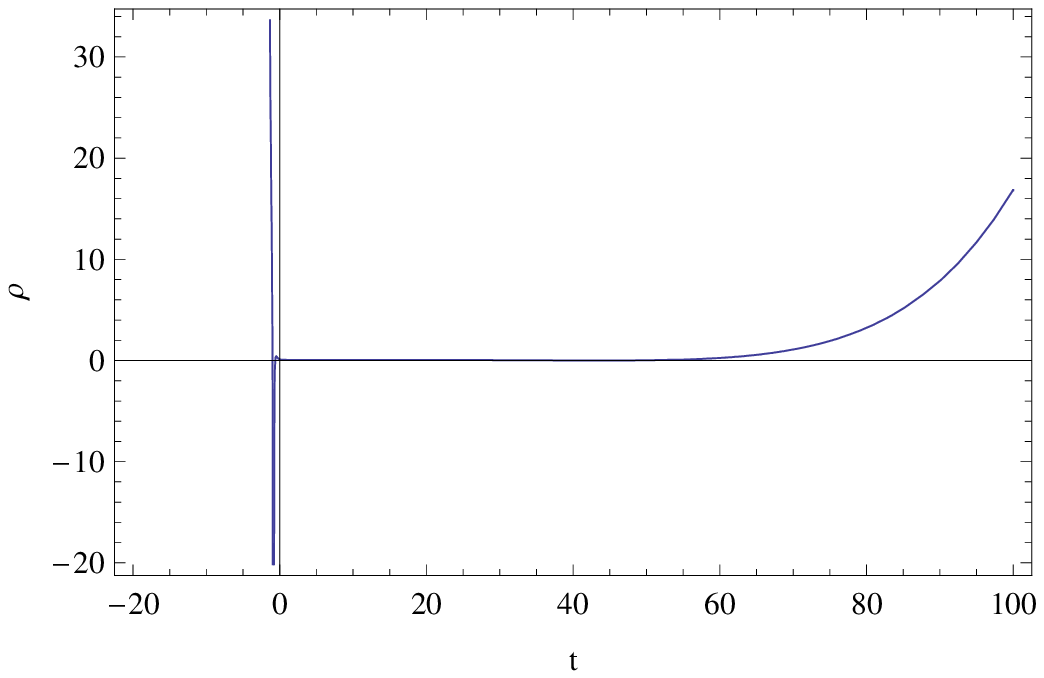}~~~~
\includegraphics[height=2.0in]{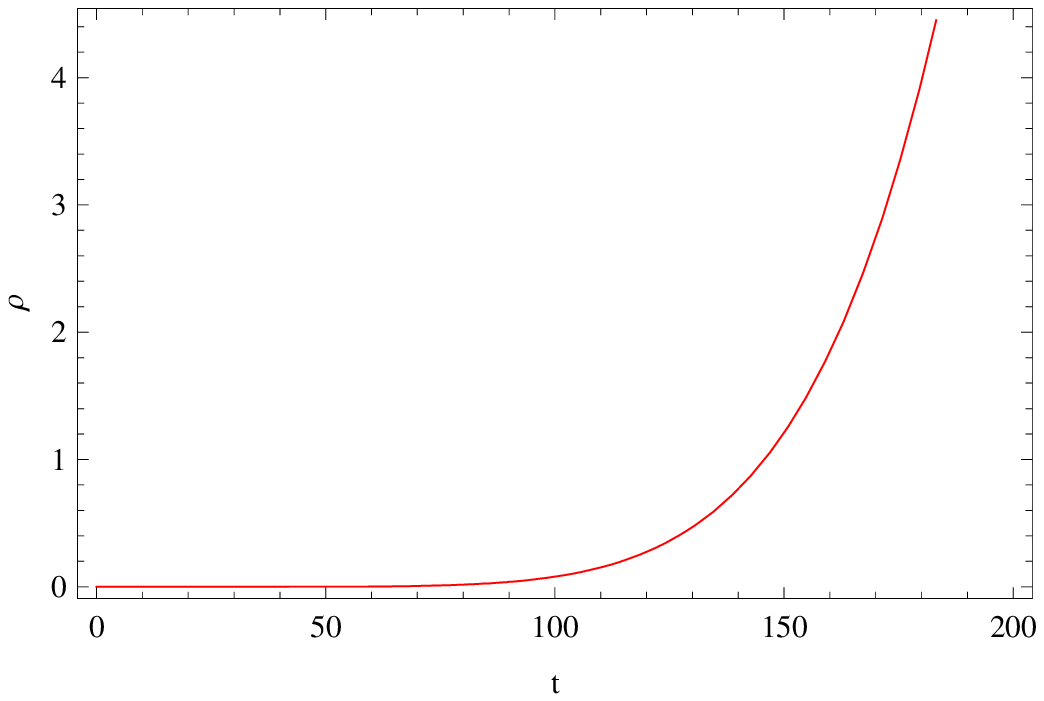}\\
\vspace{1mm}

\vspace{2mm} ~~~~~~~~~~~~Fig.7~~~~~~~~~~~~~~~~~~~~~~~~~~~~~~~~~~~~~~~~~~~Fig.8~~~~\\
\vspace{4mm}

Figs. 1 and 2 show the dynamical behavior of the system for GCCG
(left image/blue line) and VMCG (right image/red line) models
around the critical point where the parameters are chosen as $C=1$
and $\omega=-0.5$ (Case 1). Also the time evolution the scale
factor, Hubble parameter $H$, the energy density of the GCCG
and VMCG are shown in figs. 3-8 respectively for this case.\\

\vspace{6mm}

\end{figure}

\subsection{\normalsize\bf{Case 2: $k=0,~~\alpha=1/2$}}

In GCCG model, we get stable centers assuming $\omega=-0.5$ and
$C=1$. Similar solution appears in VMCG model as well when we
choose $A=0, n=2$ and $B_0=0.5$. For GCCG and VMCG model, the
graphs of $a$ vs $H$ are shown in Figs. 9 and 10.\\

\begin{figure}
\includegraphics[height=2.0in]{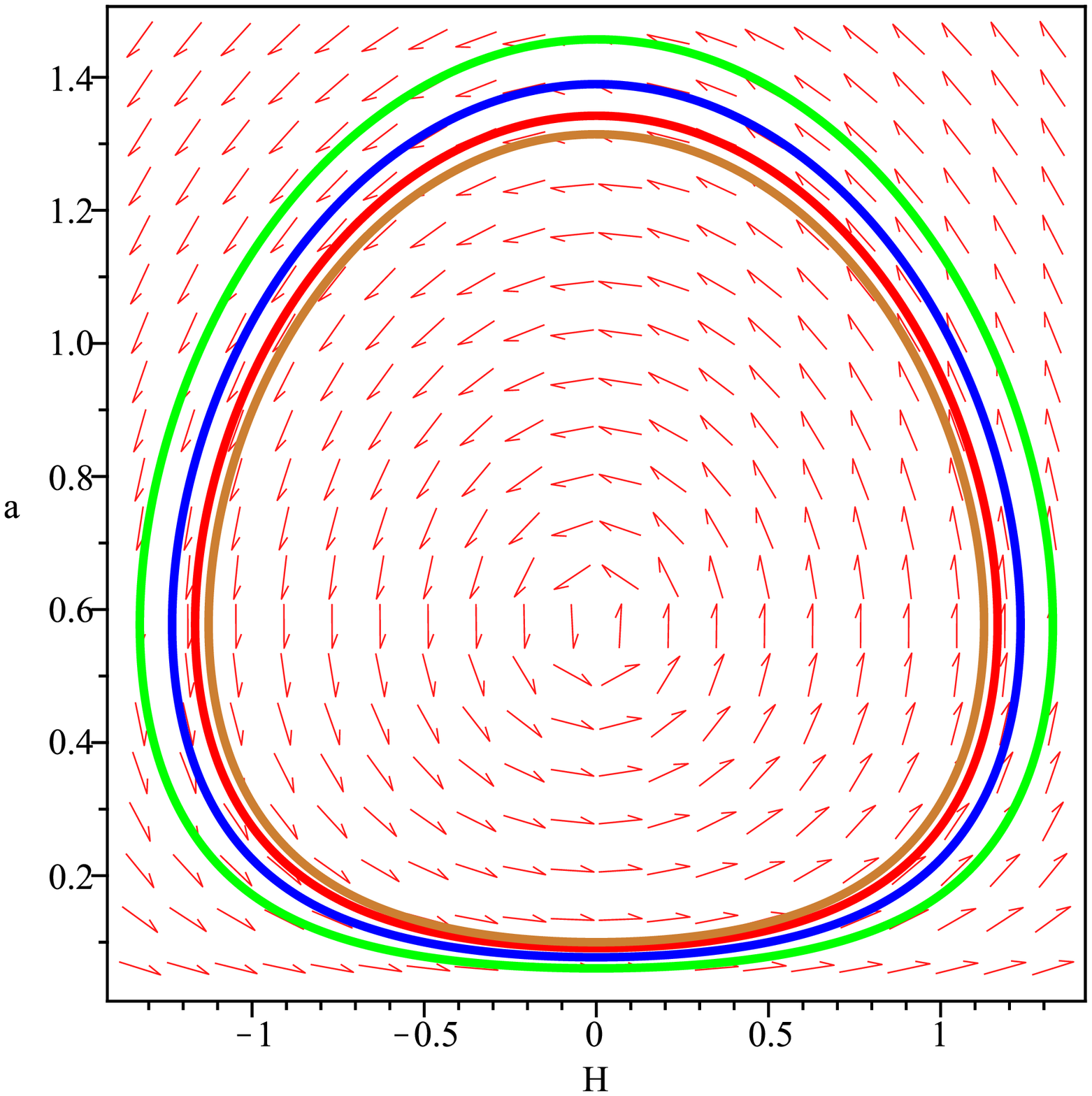}~~~~
\includegraphics[height=2.0in]{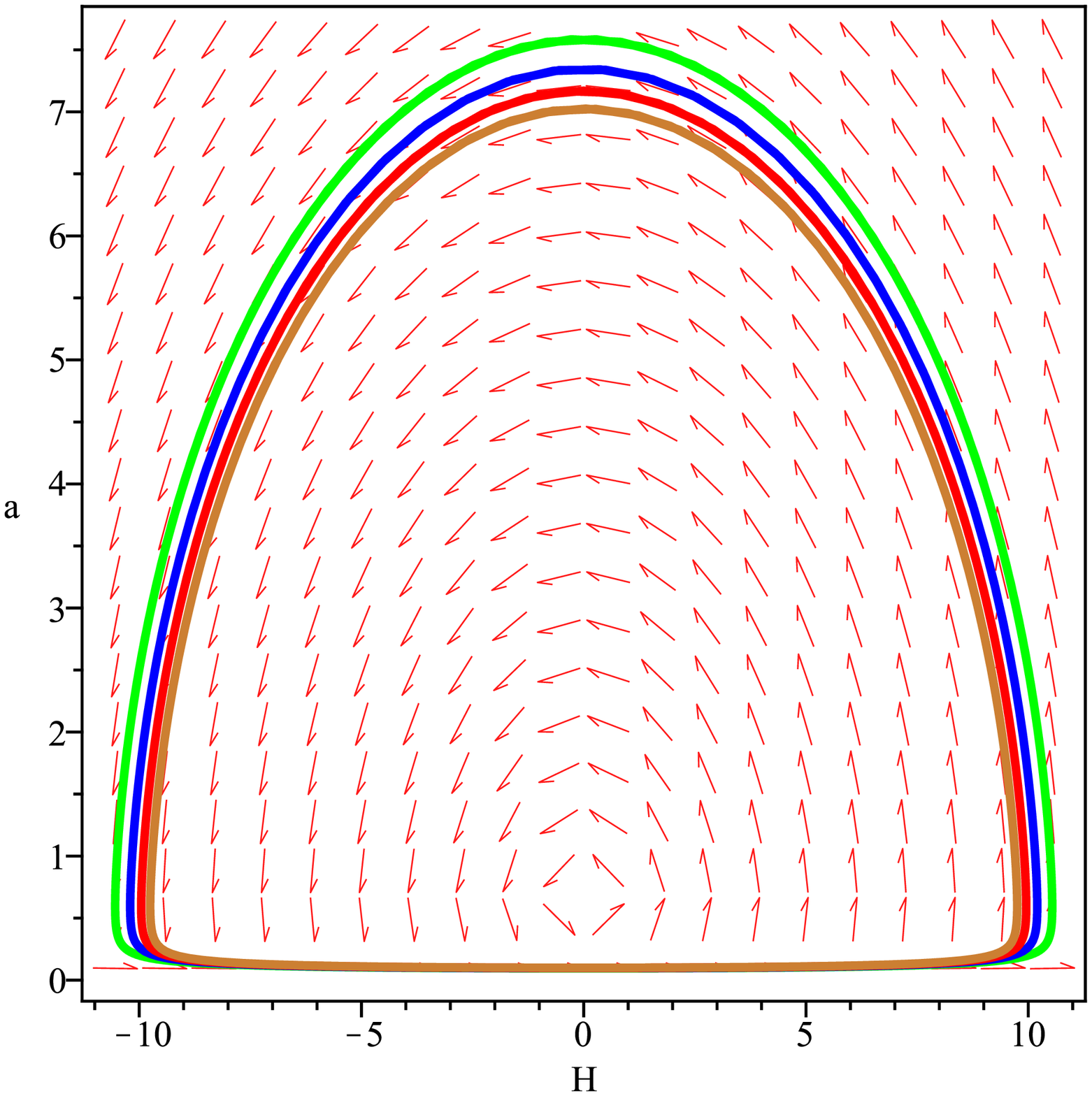}\\

\vspace{2mm} ~~~~~~~~~~~~Fig.9~~~~~~~~~~~~~~~~~~~~~~~~~~~~~~~~~~~~~~~~~~~Fig.10~~~~\\
\vspace{4mm}

Figs. 9 and 10 show the stable centers which have been found for
both GCCG (left image) and VMCG (right image) models around the
critical point where the parameters are chosen as $C=1$ and $\omega=-0.5$ (Case 2).\\

\vspace{6mm}

\end{figure}

\subsection{\normalsize\bf{Case 3: $k=1$}}

For GCCG and VMCG model, the graphs of $a$ vs $H$ are shown in
Figs. 11 and 12. Stable centers could not be obtained in this
case. If for some values of parameters, we get real eigenvalues in
both the models, then the universe does not oscillate. But a
single bounce can occur at time $t_b$ under the condition
$\dot{a}_b=0$, ${\ddot{a}_b\geq 0}$ or $\chi_b=0$ and
$\frac{d\chi_b}{dt}>0$ in terms of new variables. This gives us
the conditions for the two models from equations (10) and (11) as

\begin{equation}
C+\left[\left(\frac{3k}{a_{b}^2}\right)^{1+\alpha}-C\right]^{-\omega}>\frac{1}{3}
\left(\frac{3k}{a_{b}^2}\right)^{2+\alpha}
\end{equation}

and

\begin{equation}
B_0>\frac{3^{\alpha}k^{\alpha+1}(1+3A)}{{a_b}^{(2\alpha-n)+2}}
\end{equation}

with the energy density at the bounce for both cases as

\begin{equation}\label{bounce density}
\rho_b=\frac{3k}{a_{b}^2}
\end{equation}

To avoid the negative energy density, $\rho_b$ must be
positive, which is evident from equation \eqref{bounce density}.\\

\begin{figure}
\includegraphics[height=2.0in]{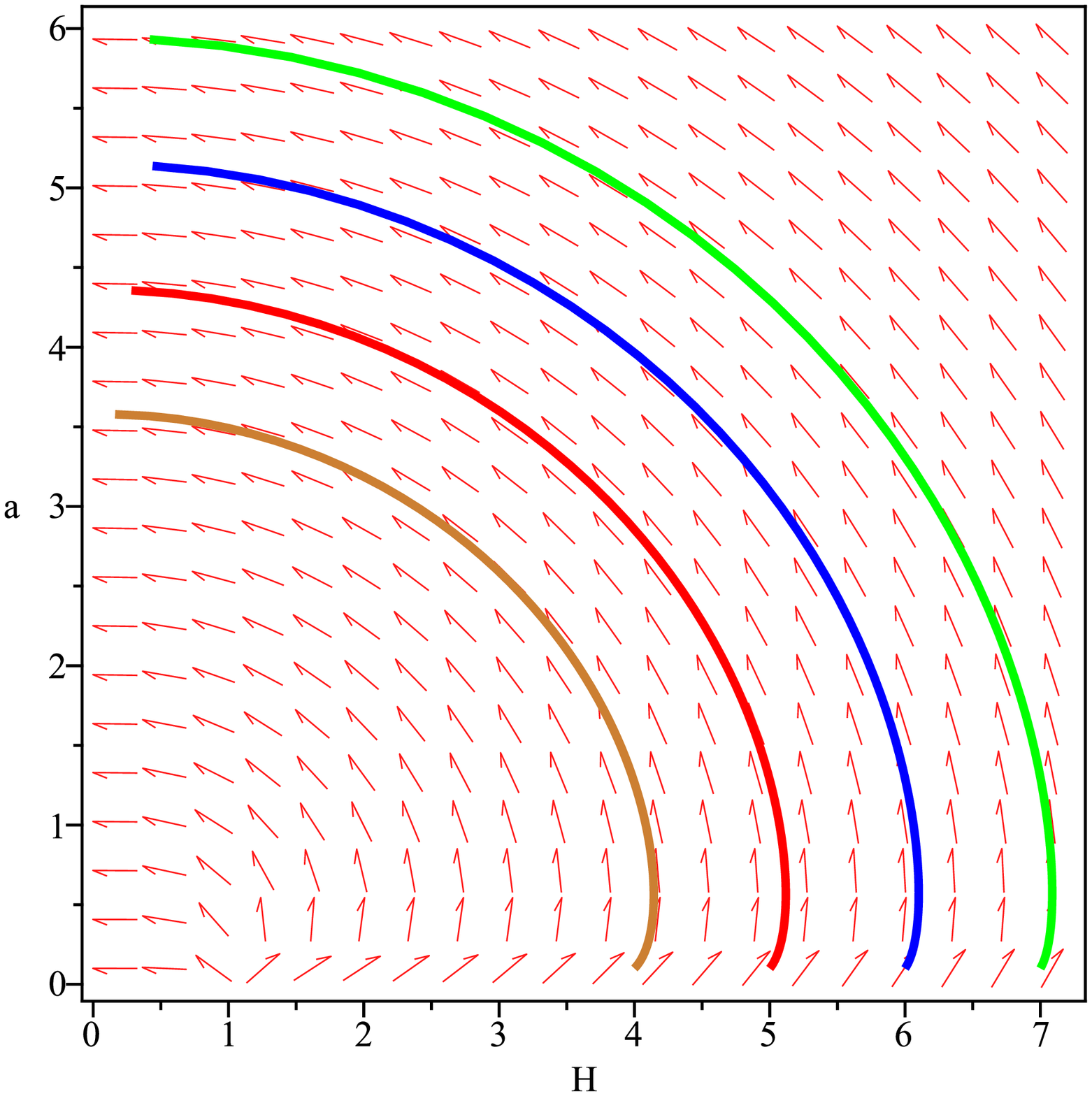}~~~~
\includegraphics[height=2.0in]{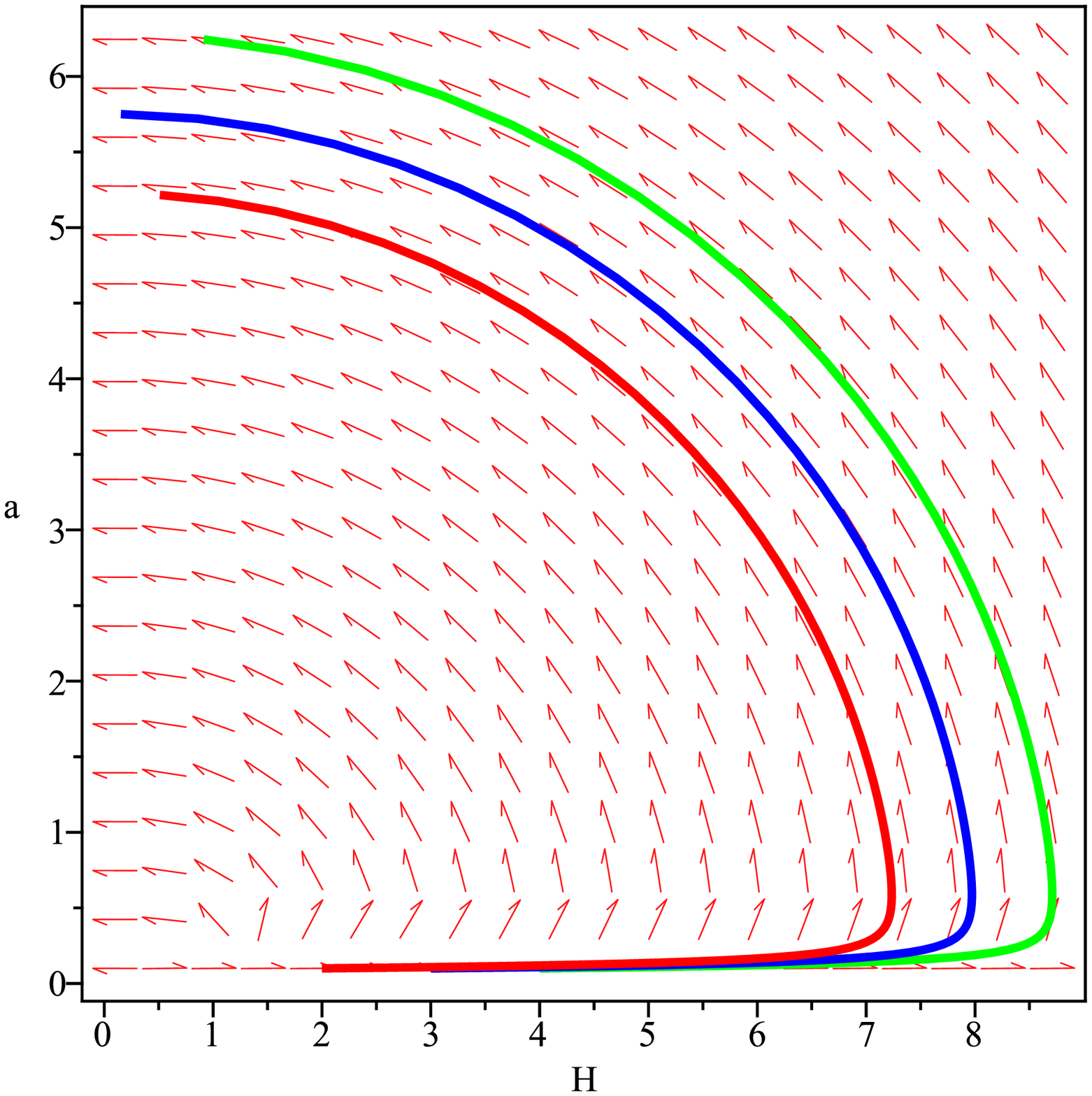}\\

\vspace{2mm} ~~~~~~~~~~~~Fig.11~~~~~~~~~~~~~~~~~~~~~~~~~~~~~~~~~~~~~~~~~~~Fig.12~~~~\\
\vspace{4mm}

Figs. 11 and 12 show the  GCCG (left image) and VMCG (right image)
models around the critical point where the parameters are
chosen as $C=1$ and $\omega=-0.5$.\\

\vspace{6mm}

\end{figure}

\section{\normalsize\bf{Bouncing Universe in Presence of Scalar Field}}

With the effective Lagrangian
$L_{\phi}=\frac{1}{2}{\dot{\phi}}^2-V$, the field equations for
both the cases are (assuming $8\pi G=1$)

\begin{equation}\label{fld eqn scalar fld 1}
3\left(H^2+\frac{k}{a^2}\right)=\rho+\frac{1}{2}{\dot{\phi}}^2+V
\end{equation}

\begin{equation}\label{fld eqn scalar fld 2}
\left(2\dot{H}+3H^2+\frac{k}{a^2}\right)=-p-\frac{1}{2}{\dot{\phi}}^2+V
\end{equation}

The energy conservation equation is given ny

\begin{equation}
\dot{\rho}+3H(\rho+p)=0
\end{equation}

The field equation for the scalar field is

\begin{equation}
\ddot{\phi}+3H\dot{\phi}+\frac{dV}{d\phi}=0
\end{equation}

Let us define the new variables

\begin{equation}
x=H,~~~y=\rho,~~~\dot{\phi}=g(t),~~~z=V
\end{equation}

Then we get the new set of equations of motion as

\begin{equation}
\frac{dx}{dt}|_{_{GCCG}}=-\frac{f_{1}(y)}{2}-\frac{1}{2}g(t)^{2}+\frac{k}{a^2},
~~~\frac{dx}{dt}|_{_{VMCG}}=-\frac{f_{2}(y)}{2a^n}-\frac{1}{2}g(t)^{2}+\frac{k}{a^2}
\end{equation}

\begin{equation}
\frac{dy}{dt}|_{_{GCCG}}=-3xf_{1}(y),~~~\frac{dy}{dt}|_{_{VMCG}}=-3x\frac{f_{2}(y)}{a^n}
\end{equation}

\begin{equation}
\frac{da}{dt}=ax
\end{equation}

\begin{equation}
\frac{dz}{dt}=-g(t)[\dot{g(t)}+3xg(t)]
\end{equation}

where
$f_{1}(y)=(\rho_{_{GCCG}}+p_{_{GCCG}})=\frac{(y^{1+\alpha}-C)-(y^{1+\alpha}-C)^{-\omega}}{y^{\alpha}}$
and $f_{2}(y)=(\rho_{_{VMCG}}+p_{_{VMCG}})=\frac{(1+A)y^{1+\alpha}-B_0}{y^{\alpha}}$.\\

For $g(t)=0$ in both the cases, the scale factor is seen to have a
single bounce without any oscillations. Figs. 13-15 show that
dynamical behavior of the scale factor and the Hubble parameter in
GCCG and VMCG models.\\

\begin{figure}
\includegraphics[height=2.0in]{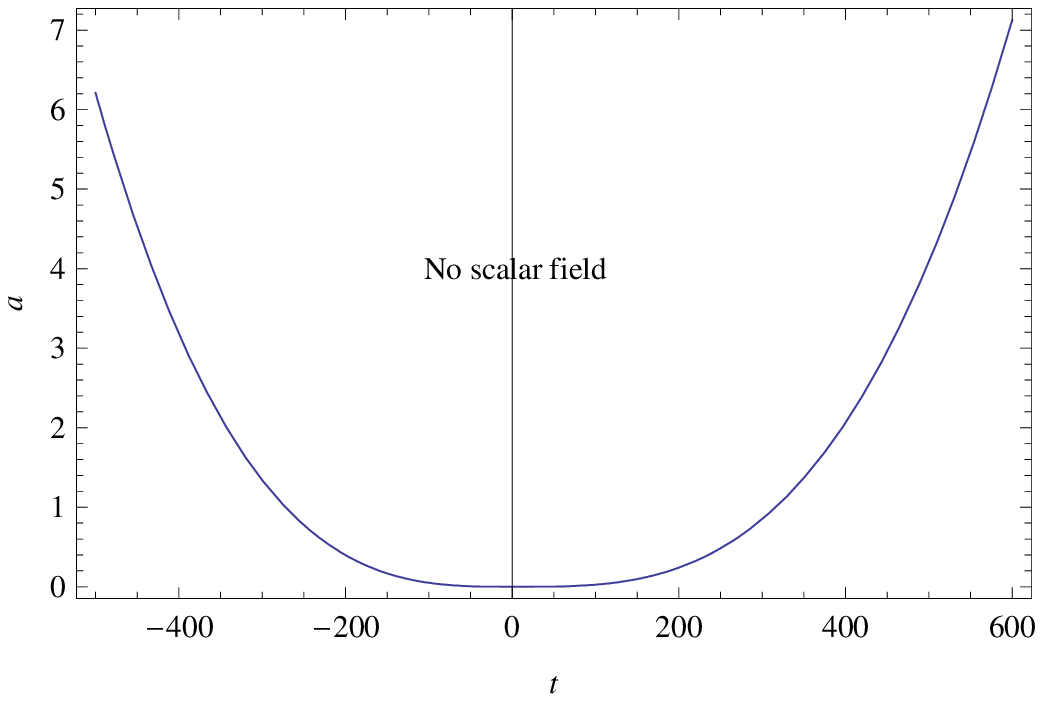}~~~~
\includegraphics[height=2.0in]{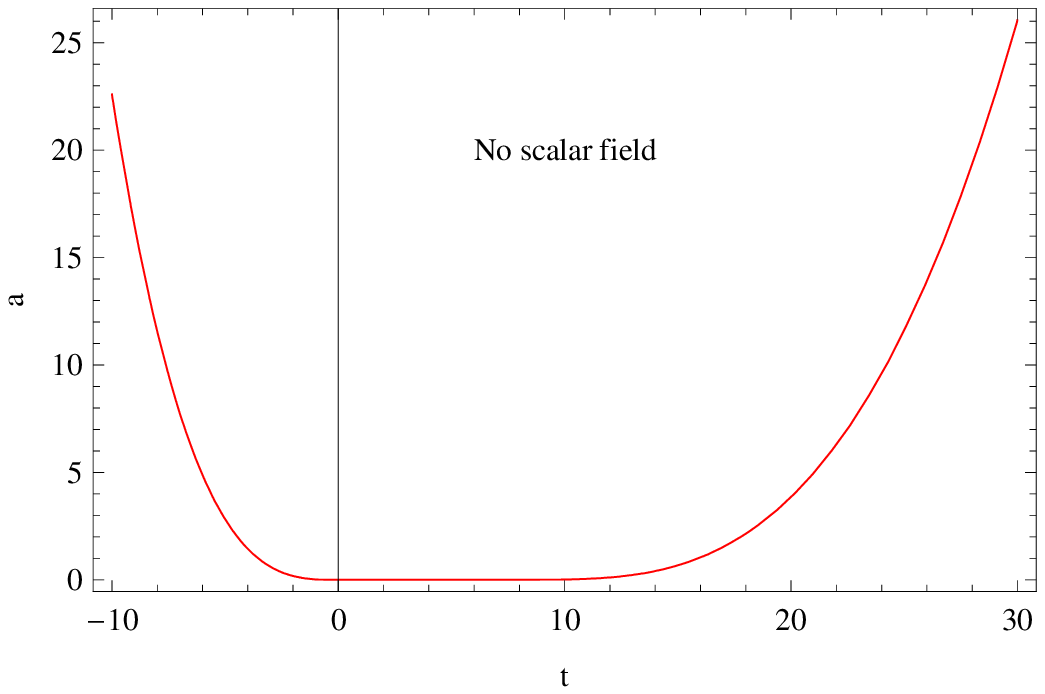}\\

\vspace{2mm} ~~~~~~~~~~~~Fig.13~~~~~~~~~~~~~~~~~~~~~~~~~~~~~~~~~~~~~~~~~~~Fig.14~~~~\\
\vspace{4mm}

\includegraphics[height=2.0in]{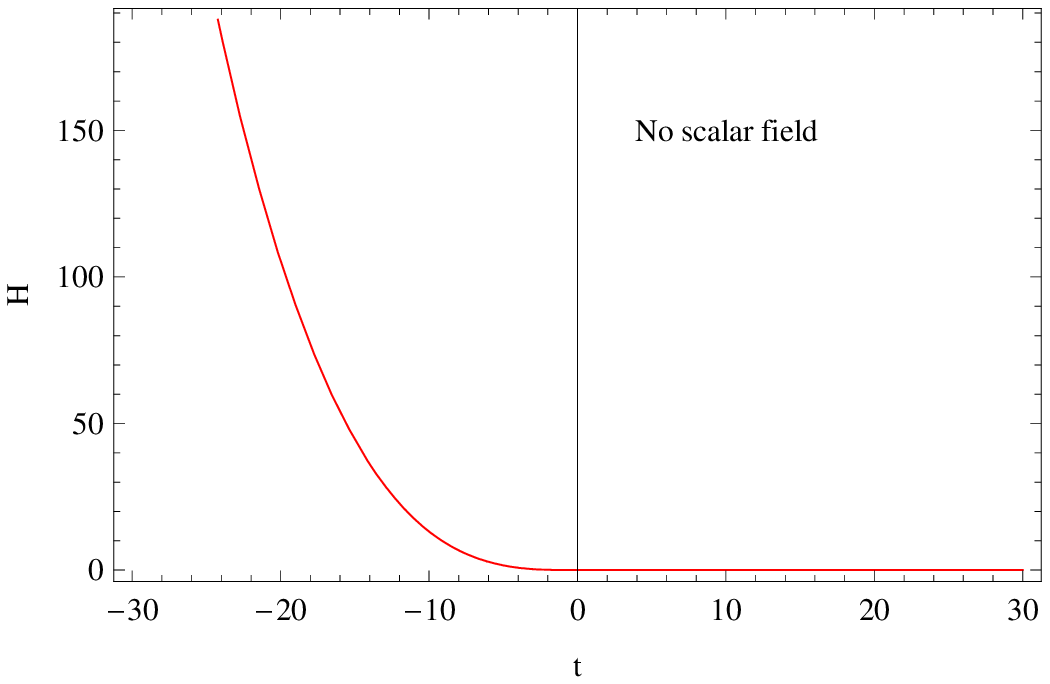}\\
\vspace{2mm} ~~~~~~~~~~~~~~~~~~~~~~~~Fig.15~~~~~~~~~~~~~~\\

\vspace{4mm} Figs. 13-15 show the dynamical behavior of the scale
factor and the Hubble parameter in GCCG (blue line/Fig.13) and
VMCG (red line/Figs.14 and 15) models for $k=0$ and $g(t)=0$.
Figs. 13 and 14 signify that in the absence of the scalar field in
a flat universe, the models undergo a single bounce only and
do not show any oscillations.\\

\vspace{6mm}

\end{figure}

For the second choice of $g(t)=e^{-\lambda t}$, ($\lambda$
constant, damping coefficient), we have the minimal conditions of
bounce as $\chi_b=0,~~\frac{d\chi_b}{dt}>0$, which in terms of the
new parameters for the two models become

\begin{equation}
C+({y_b}^{1+\alpha}-C)^{-\omega}>\left[y_b+{g_b}^2-\frac{2k}{a_{b}^2}\right]{y_b}^{\alpha}
\end{equation}

and

\begin{equation}
B_0>{y_{b}}^{\alpha}\left[(1+A)y_b+\left({g_b}^2-\frac{2k}{a_{b}^2}\right){a_b}^n\right]
\end{equation}

where $g_b=e^{-\lambda t_b}$. The dynamical behavior of this case
has been shown in Figs. 16-23 with different $k$ values. In none
of these cases, the scale factor has shown either a single
bounce or an oscillating behaviour for different parameters.\\

\begin{figure}
\includegraphics[height=2.0in]{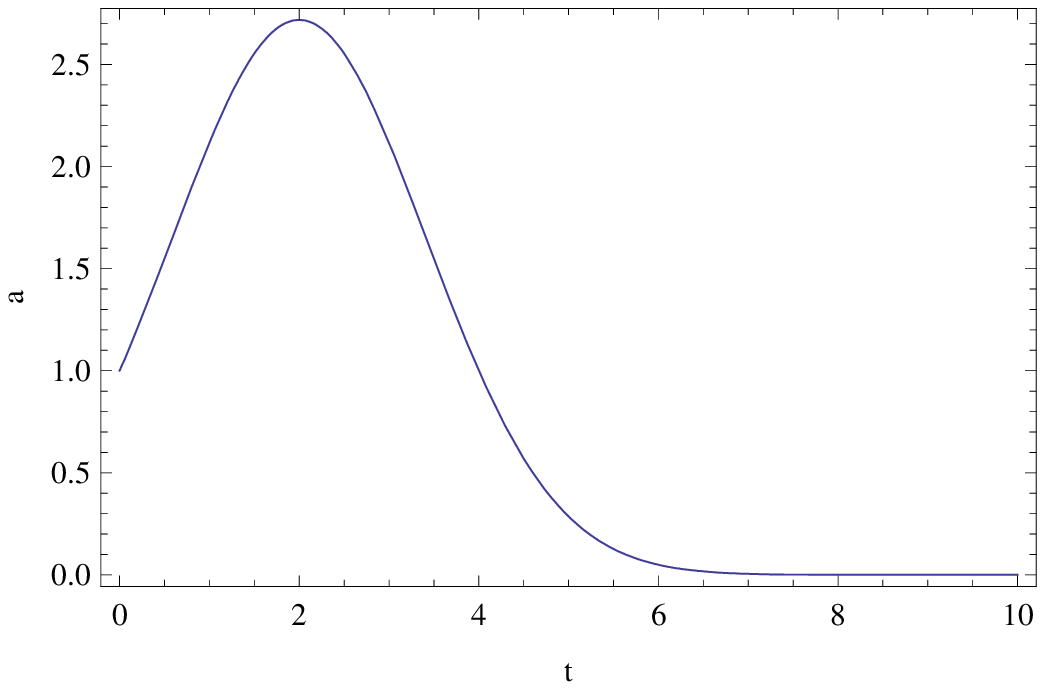}~~~~
\includegraphics[height=2.0in]{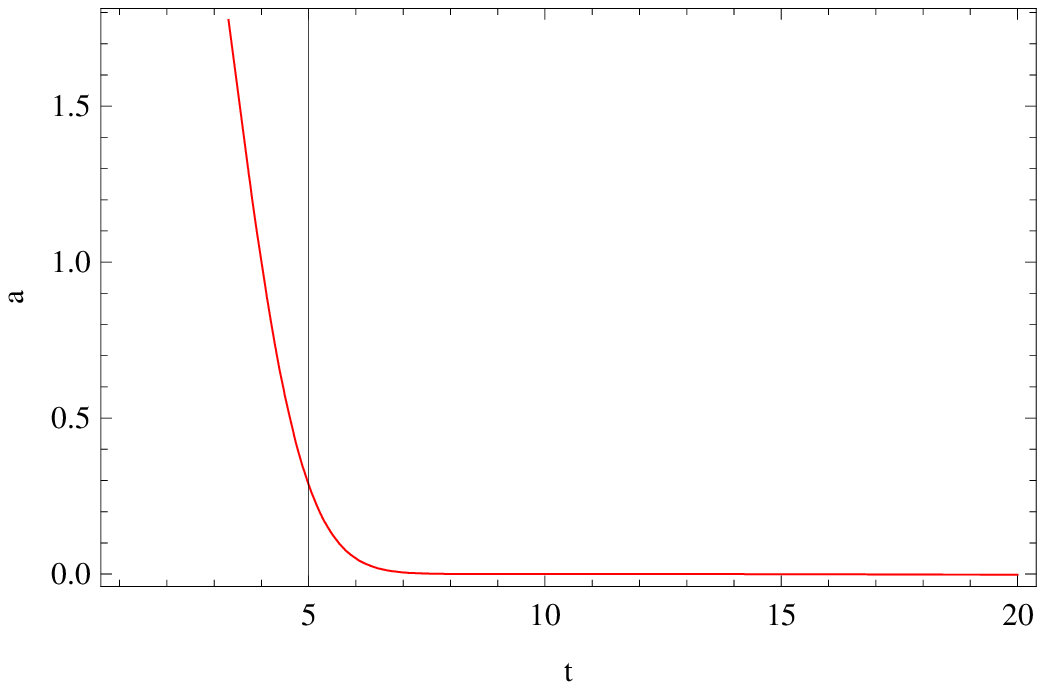}\\

\vspace{2mm} ~~~~~~~~~~~~Fig.16~~~~~~~~~~~~~~~~~~~~~~~~~~~~~~~~~~~~~~~~~~~Fig.17~~~~\\
\vspace{4mm}

\includegraphics[height=2.0in]{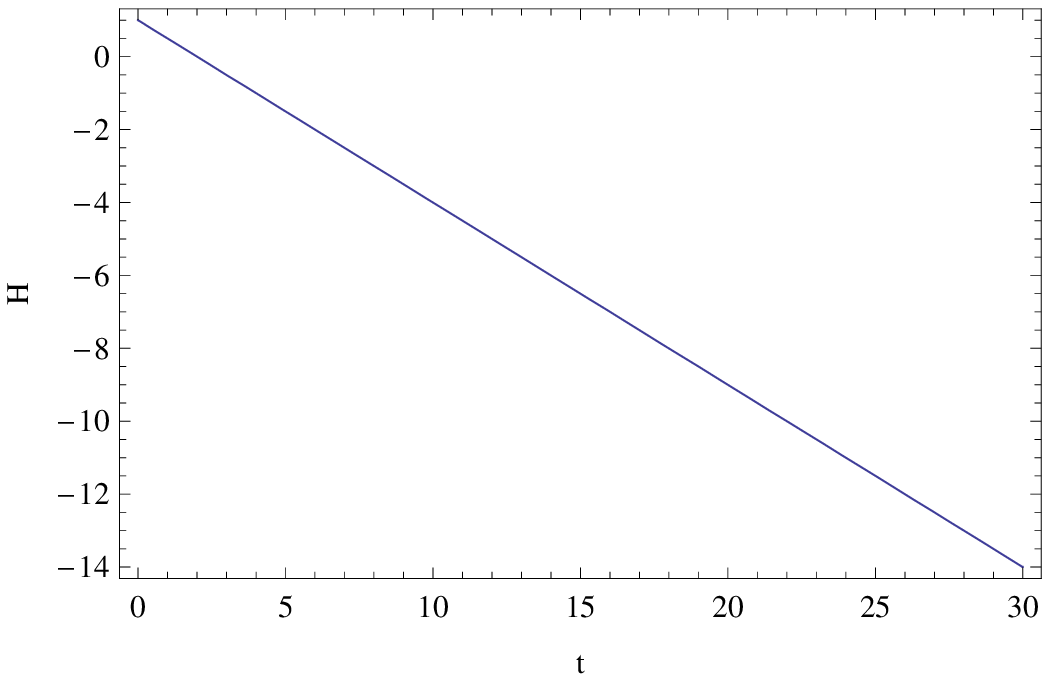}~~~~
\includegraphics[height=2.0in]{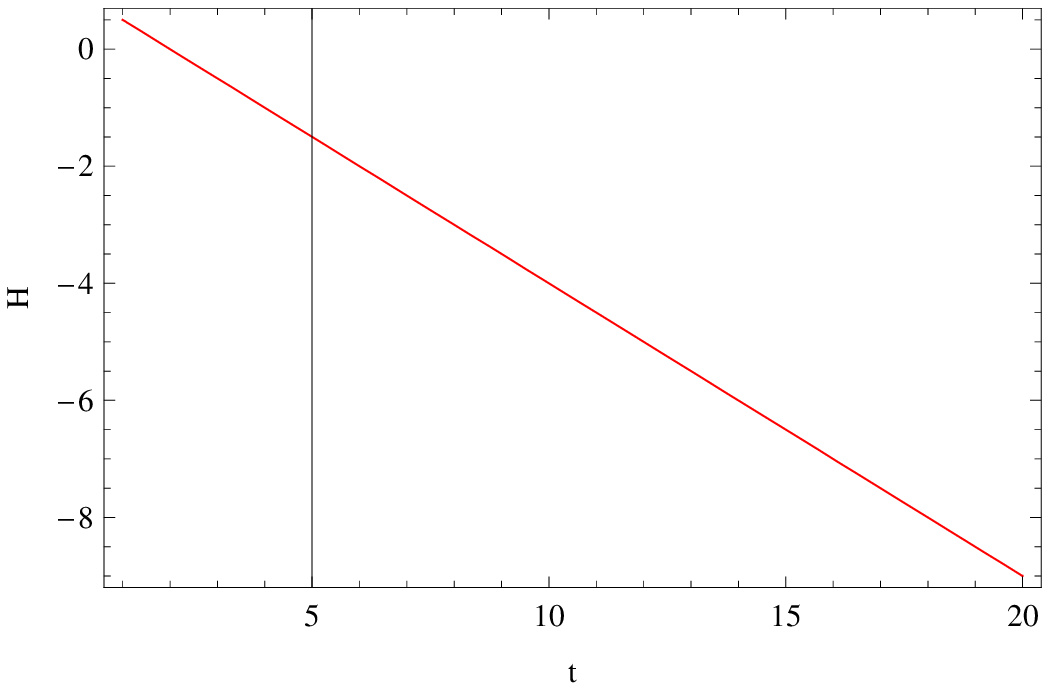}\\

\vspace{2mm} ~~~~~~~~~~~~Fig.18~~~~~~~~~~~~~~~~~~~~~~~~~~~~~~~~~~~~~~~~~~~Fig.19~~~~\\
\vspace{4mm}

Figs. 16-19 show the dynamical behavior of the scale factor and
the Hubble parameter in GCCG (blue line/left images) and VMCG (red
line/right images) models for $k=0$ and $g(t)=e^{-\lambda t}$.\\

\vspace{4mm}

\end{figure}

\begin{figure}

\includegraphics[height=2.0in]{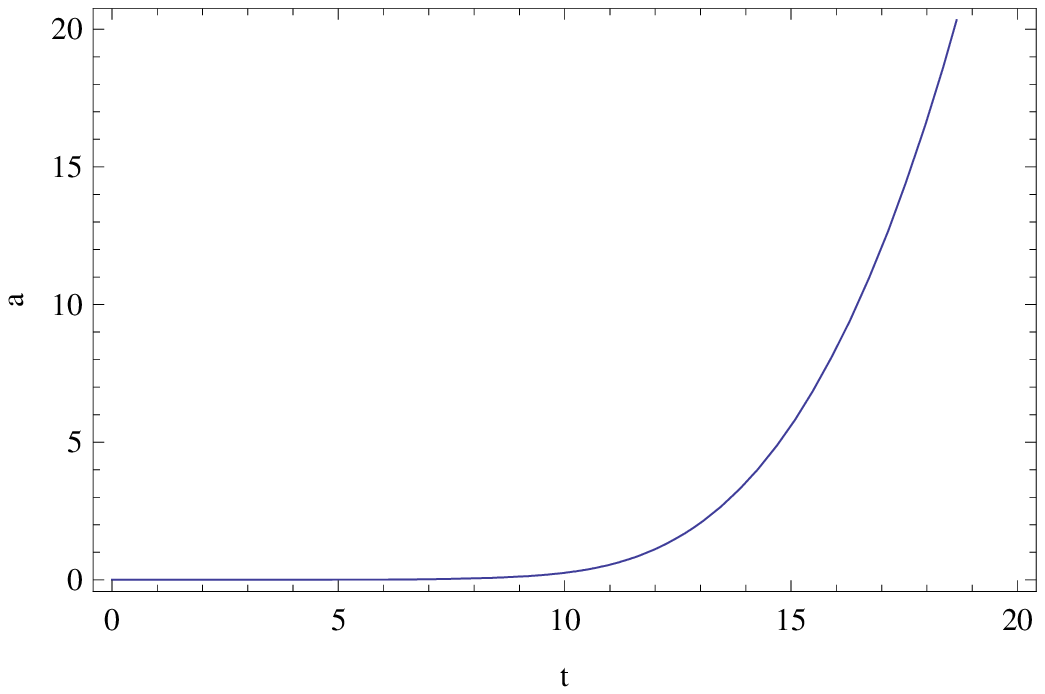}~~~~
\includegraphics[height=2.0in]{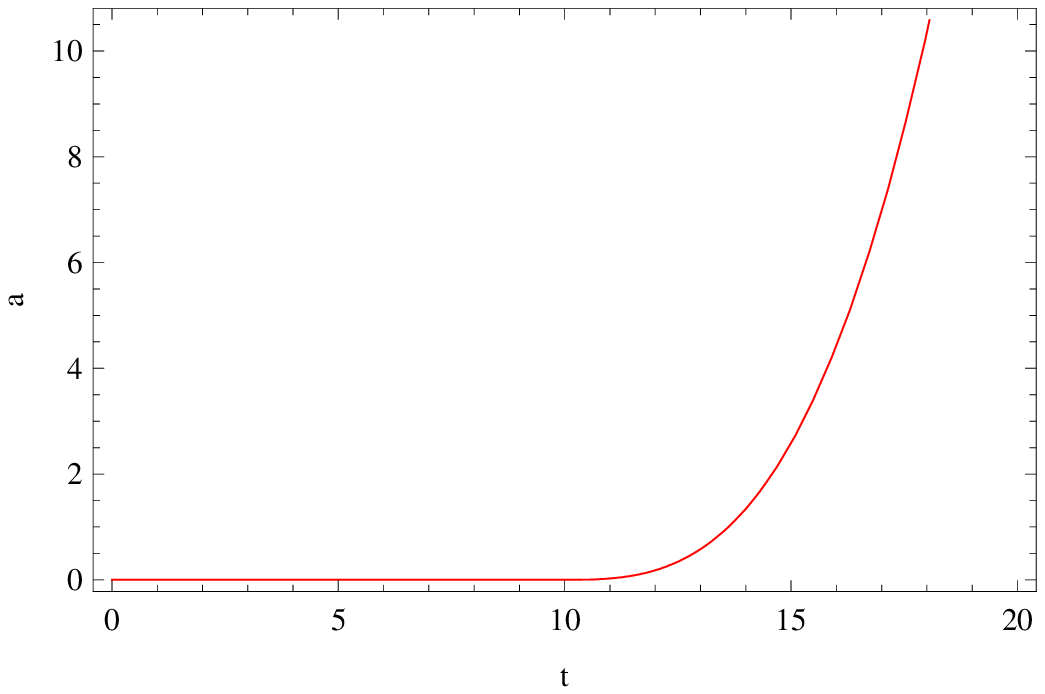}\\

\vspace{2mm} ~~~~~~~~~~~~Fig.20~~~~~~~~~~~~~~~~~~~~~~~~~~~~~~~~~~~~~~~~~~~Fig.21~~~~\\
\vspace{4mm}

Figs. 20 and 21 show the dynamical behavior of the scale factor in
GCCG (blue line/left image) and VMCG (red line/right image) models
for $k=1$ and $g(t)=e^{-\lambda t}$.\\

\vspace{4mm}

\includegraphics[height=2.0in]{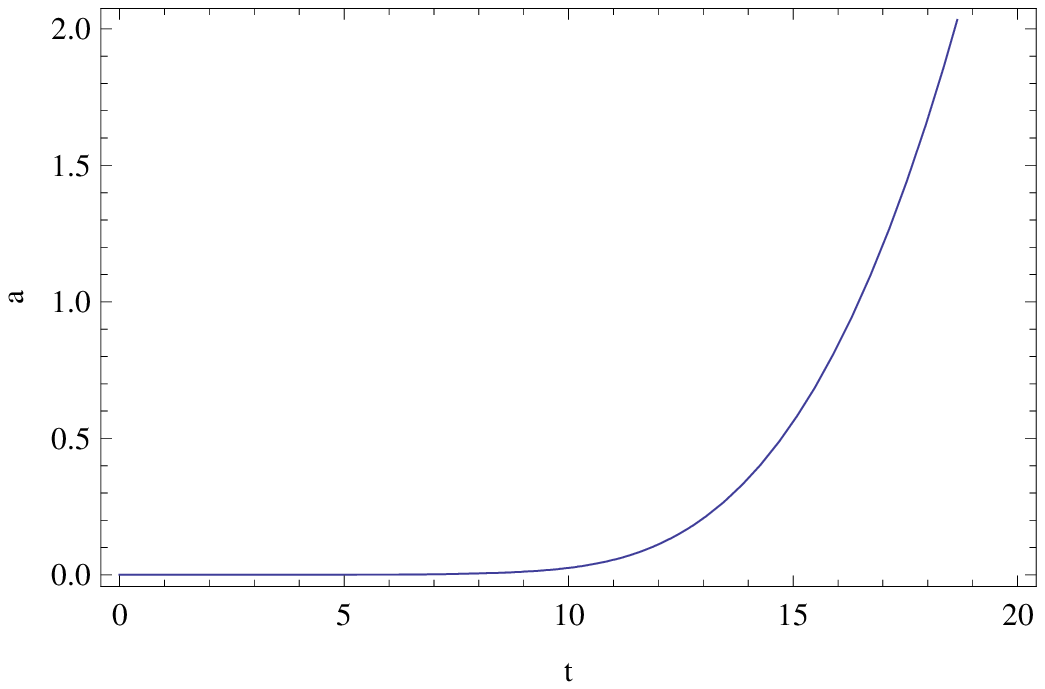}~~~~
\includegraphics[height=2.0in]{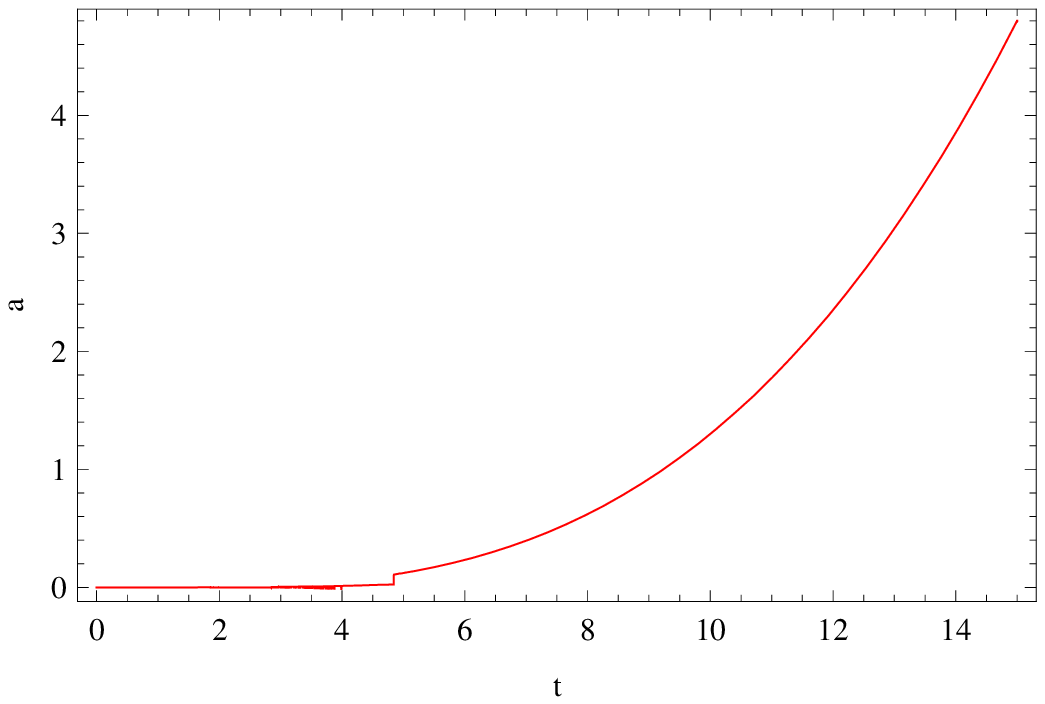}\\

\vspace{2mm} ~~~~~~~~~~~~Fig.22~~~~~~~~~~~~~~~~~~~~~~~~~~~~~~~~~~~~~~~~~~~Fig.23~~~~\\
\vspace{4mm}

Figs. 22 and 23 show the dynamical behavior of the scale factor in
GCCG (blue line/left image) and VMCG (red line/right image) models
for $k=-1$ and $g(t)=e^{-\lambda t}$.\\

\vspace{4mm}

\end{figure}

\section{\normalsize\bf{Connection between Bounce and Energy Conditions}}

In this section, we demonstrate the connection between bounce and
energy conditions. Reframing the field equations, we have

\begin{equation}\label{new fld eqn 1}
\rho+p=2\left[\frac{k}{a^2}+\frac{\dot{a}^2}{a^2}-\frac{\ddot{a}}{a}\right]
\end{equation}

\begin{equation}\label{new fld eqn 2}
\rho-p=2\left[\frac{2k}{a^2}+2\frac{\dot{a}^2}{a^2}+\frac{\ddot{a}}{a}\right]
\end{equation}

\begin{equation}
\rho+3p=-6\frac{\ddot{a}}{a}
\end{equation}

so that for bounce, the restrictions become

\begin{equation}\label{restriction 1}
\rho_b+p_b<\frac{2k}{a_{b}^2}
\end{equation}

\begin{equation}
\rho_b-p_b>\frac{4k}{a_{b}^2}
\end{equation}

\begin{equation}
\rho_b+3p_b<0
\end{equation}

In the discussion of energy conditions, we mainly focus on NEC
$\Leftrightarrow (\rho+p\geq 0)$. If this is violated, then all
the other point-wise energy conditions are
violated as well.\\\\

\subsection{\normalsize\bf{NEC in spatially flat ($k=0$) and hyperbolic/open ($k=-1$) universe}}

From equation \eqref{restriction 1}, NEC is definitely violated.
Let us see the energy conservation equation for the bounce

\begin{equation}\label{bounce continuity eqn}
\dot{\rho_b}=-3H_b(\rho_b+p_b)
\end{equation}

Since $H_b=0$, hence $\dot{\rho_b}=0$. Thus the energy densities
of both the models reach a point of extremum at the time of
bounce. Now

\begin{equation}\label{ddt bounce density}
\ddot{\rho_b}=-3\dot{H_b}(\rho_b+p_b)-3H_b(\dot{\rho_b}+\dot{p_b})=-3\dot{H_b}(\rho_b+p_b)
\end{equation}

Now, for bounce $\ddot{a_b}>0$, which implies $\dot{H_b}>0$. Also
restriction \eqref{restriction 1} gives violation of NEC. Hence
from equation \eqref{ddt bounce density}, we must have,
$\ddot{\rho_b}>0~\Rightarrow
\rho_b=\rho_{min}=\frac{3k}{a_{b}^2}$.\\

Then from the equations of state \eqref{EoS GCCG} and \eqref{EoS
VMCG}, the restrictions for the NEC violation with bounce for GCCG
and VMCG models will be

\begin{equation}
C+\left(\frac{3k}{a_{b}^2}-C\right)^{-\omega}>\left(\frac{3k}{a_{b}^2}\right)^{\alpha+1}
\end{equation}

and

\begin{equation}
\frac{3k}{{a_b}^{\frac{(2\alpha-n)+2}{\alpha+1}}}-\left(\frac{B_0}{A+1}\right)^{\frac{1}{\alpha+1}}<0
\end{equation}

\subsection{\normalsize\bf{NEC in hyperspherical/closed ($k=1$) universe}}

Here we do not get violation of NEC automatically. For bounce,
here we rewrite equation \eqref{new fld eqn 1} as follows

\begin{equation}
\rho_b+p_b=-2\left[\frac{\ddot{a_b}}{a_b}-\frac{1}{a_{b}^2}\right]
\end{equation}

This follows that NEC will be violated if
$\ddot{a_b}\geq\frac{1}{a_b}$. Thus there are two possibilities\\

(a) $\ddot{a_b}\geq\frac{1}{a_b}$. This implies $\dot{H_b}>0$ and
$(\rho_b+p_b)<0$. Hence we must have $\ddot{\rho_b}>0 \Rightarrow
\rho_b=\rho_{min}=\frac{3}{a_{b}^2}$.\\

(b) $\ddot{a_b}<\frac{1}{a_b}$. This again implies $\dot{H_b}>0$
but this time NEC is satisfied, i.e, $(\rho_b+p_b)>0$. Therefore,
$\ddot{\rho_b}<0 \Rightarrow
\rho_b=\rho_{max}=\frac{3}{a_{b}^2}$.\\

Thus for a bounce, violation of NEC is not necessary.
Nevertheless, if NEC is violated, then the energy density reaches
its minimum value at the bounce point. Thus for the EoS \eqref{EoS
GCCG} and \eqref{EoS VMCG}, the restrictions of NEC violation
under the bounce for the GCCG and VMCG models will be two
bi-conditionals viz.

\begin{equation}
C+\left(\frac{3}{a_{b}^2}-C\right)^{-\omega}>\left(\frac{3}{a_{b}^2}\right)^{\alpha+1}~\Leftrightarrow
\ddot{a_b}\geq\frac{1}{a_b}
\end{equation}

and

\begin{equation}
B_0>\frac{3^{\alpha}(A+1)}{{a_b}^{(2\alpha-n)+2}}
\end{equation}

\begin{figure}
\vspace{4mm}

\includegraphics[height=2.0in]{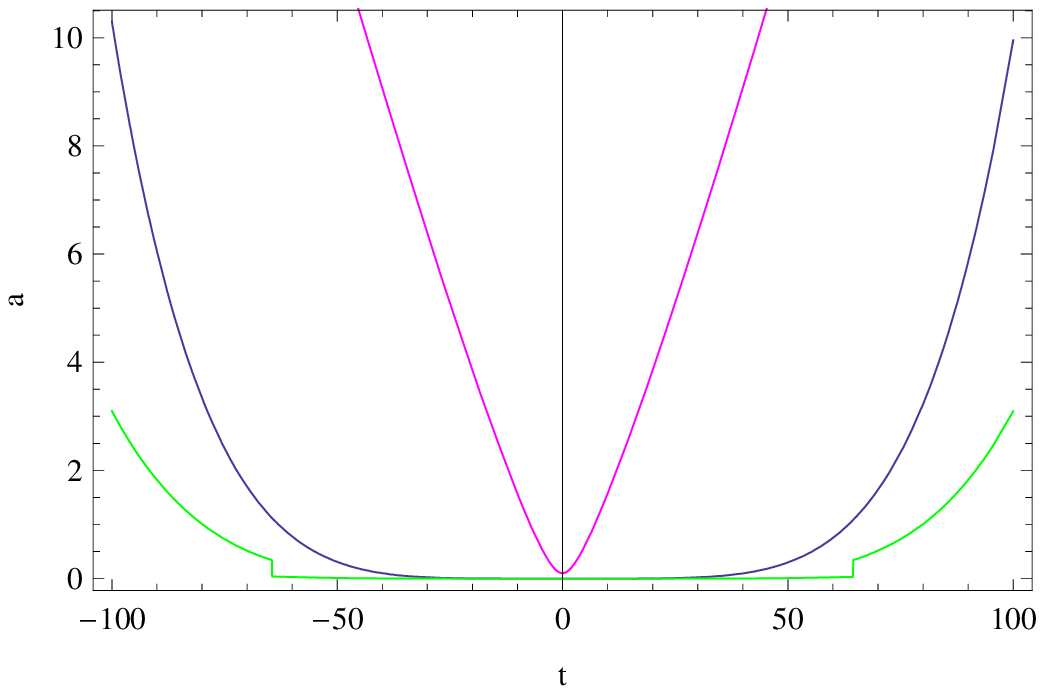}~~~~
\includegraphics[height=2.0in]{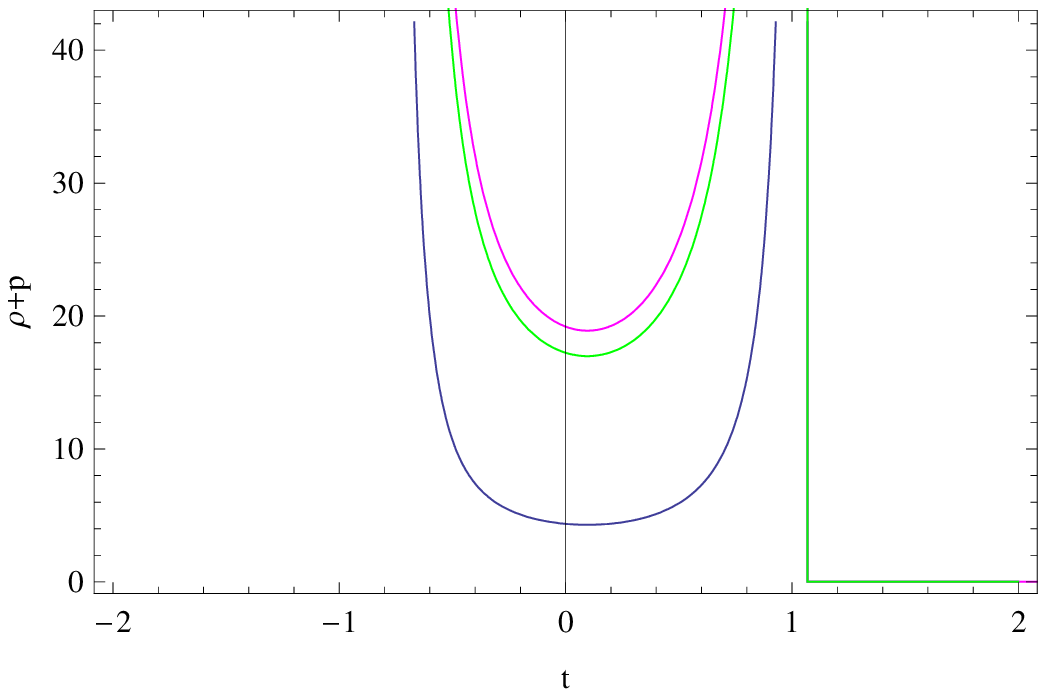}\\
\vspace{2mm} ~~~~~~~~~~~~Fig.24~~~~~~~~~~~~~~~~~~~~~~~~~~~~~~~~~~~~~~~~~~~Fig.25~~~~\\
\vspace{4mm}

\includegraphics[height=2.0in]{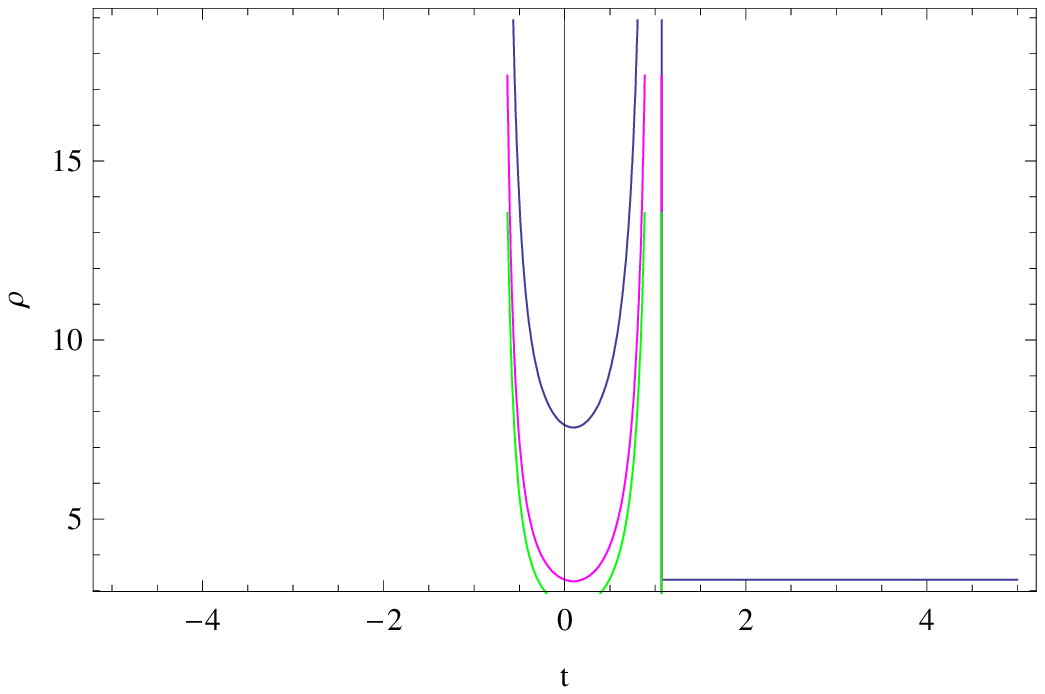}\\
\vspace{2mm} ~~~~~~~~~~~~~~~~~~~~~~~~Fig.26~~~~~~~~~~~~~~\\

Figs. 24-26 show the dynamical behavior of the scale factor $a$,
the sum of the energy density and pressure
$(\rho_{_{GCCG}}+p_{_{GCCG}})$ and the energy density
$\rho_{_{GCCG}}$ during the bounce period for $k=1$, $\alpha=0.5$,
$\omega=-0.5$ and $a_b=1$ for different values of $C$ and $D$ from
\eqref{rho gccg}. The plots show that for different values of
$C$ and $D$, a bounce always appear and the NEC is always satisfied.\\

\vspace{4mm}

\end{figure}

\begin{figure}
\vspace{4mm}

\includegraphics[height=2.0in]{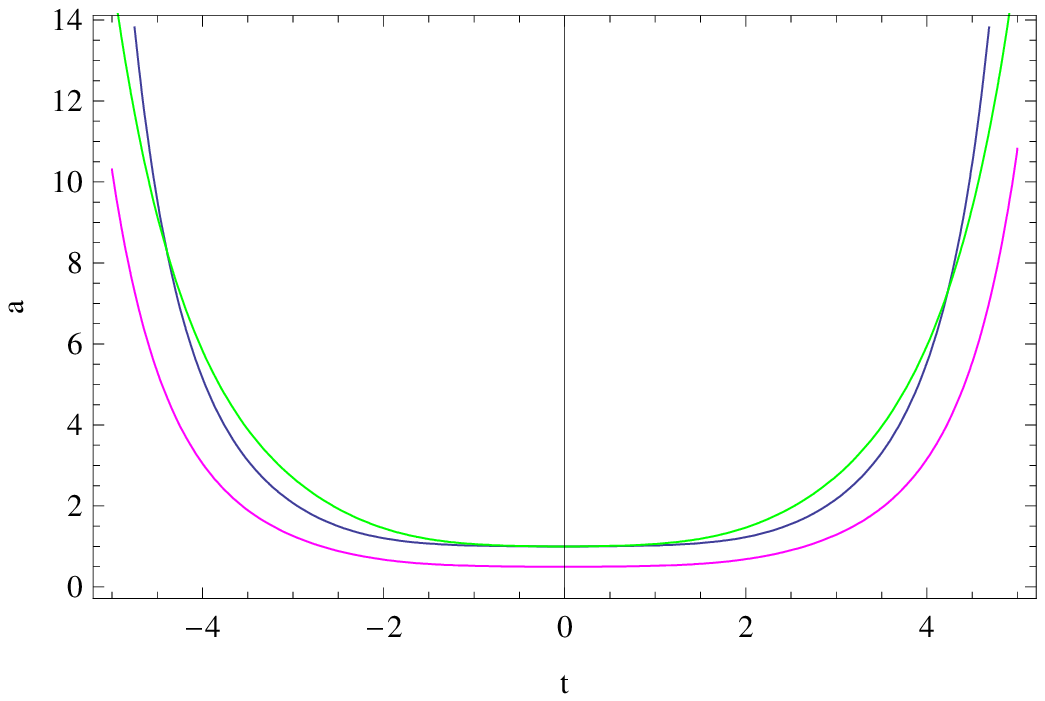}~~~~
\includegraphics[height=2.0in]{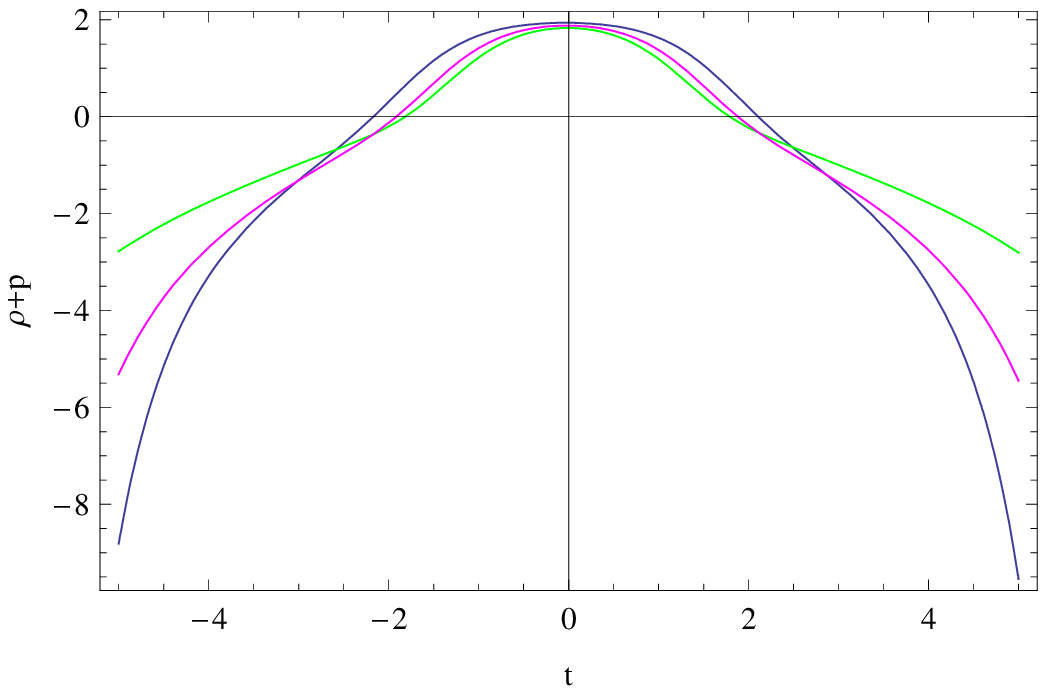}\\
\vspace{2mm} ~~~~~~~~~~~~Fig.27~~~~~~~~~~~~~~~~~~~~~~~~~~~~~~~~~~~~~~~~~~~Fig.28~~~~\\
\vspace{4mm}

\includegraphics[height=2.0in]{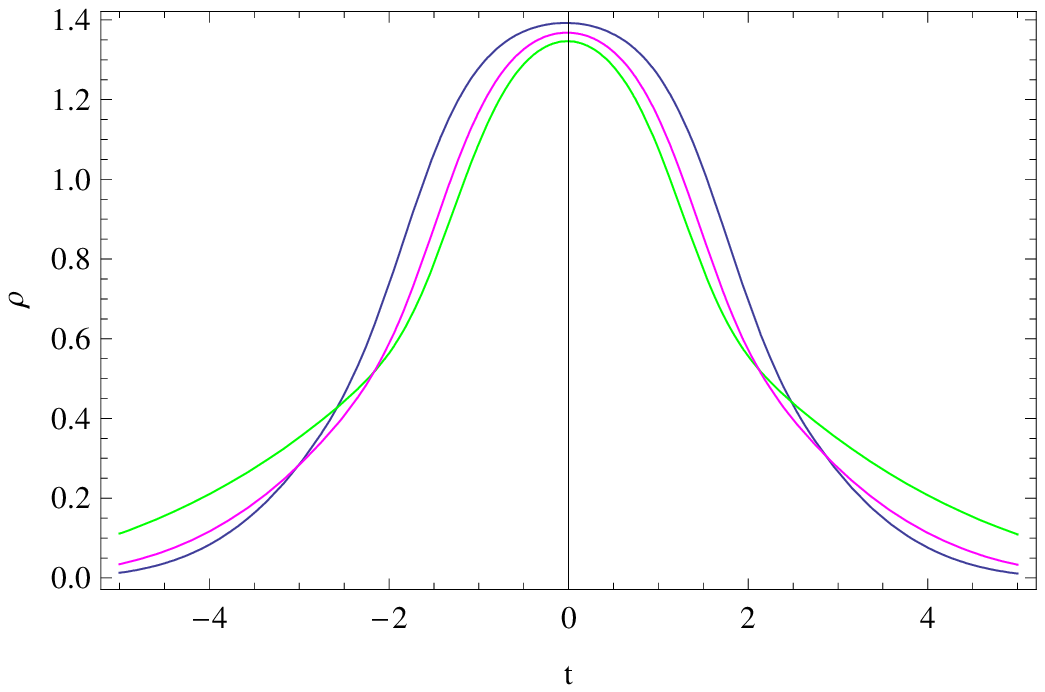}\\
\vspace{2mm} ~~~~~~~~~~~~~~~~~~~~~~~~Fig.29~~~~~~~~~~~~~~\\

Figs. 27-29 show the dynamical behavior of the scale factor $a$,
the sum of the energy density and pressure
$(\rho_{_{VMCG}}+p_{_{VMCG}})$ and the energy density
$\rho_{_{VMCG}}$ during the bounce period for $k=1$, $\alpha=0.5$,
and $a_b=1$ for different values of $n$ from \eqref{rho vmcg}. The
plots show that for all the variations in $n$, a bounce always
occur. However in this case, the NEC is violated in late
time.\\

\vspace{4mm}

\end{figure}

\underline{GCCG Model}\\

In this case, $\rho_{_{GCCG}}\rightarrow
(1+C)^{\frac{1}{1+\alpha}}$, a constant value in the future. The
Hubble parameter $H\rightarrow \pm
\frac{1}{\sqrt{3}}(1+C)^{\frac{1}{2(1+\alpha)}}$. The positive and
negative signs signifying the expanding and contracting universes
respectively. Also from the EoS \eqref{EoS GCCG}, we get
$p_{_{GCCG}}\rightarrow -(1+C)^{\frac{1}{1+\alpha}}$,
$\omega_{_{GCCG}}\rightarrow -1$ and
$(\rho_{_{GCCG}}+p_{_{GCCG}})\rightarrow 0$ in future. Therefore
the energy conservation equation \eqref{bounce continuity eqn}
gives $\frac{d\rho}{dt}\rightarrow 0$. Hence the energy density
attains a point of extremum in future as well. Therefore we have
two points ($t=t_b$ and $t\rightarrow \infty$) where the energy
density of GCCG will attain its extremum values depending on the
violation or satisfaction of the NEC. This can be classified in
the following way:\\

(a) NEC is violated $\Rightarrow
\rho_b=\rho_{min}=\frac{3}{a_{b}^2}$,
$\rho_{\infty}=\rho_{max}=(1+C)^{\frac{1}{1+\alpha}}$.\\

(b) NEC is satisfied $\Rightarrow
\rho_b=\rho_{max}=\frac{3}{a_{b}^2}$,
$\rho_{\infty}=\rho_{min}=(1+C)^{\frac{1}{1+\alpha}}$.\\

Figs. 24-26 show the dynamical behavior of the scale factor $a$,
the sum of the energy density and pressure
$(\rho_{_{GCCG}}+p_{_{GCCG}})$ and the energy density
$\rho_{_{GCCG}}$ during the bounce period for $k=1$, $\alpha=0.5$,
$\omega=-0.5$ and $a_b=1$ for different values of $C$ and $D$.\\

\underline{VMCG Model}\\

In this case, $\rho_{_{VMCG}}\rightarrow 0$ in the future. But in
that case every other quantities as $p_{_{VMCG}}$, $H$,
$(\rho_{_{VMCG}}+p_{_{VMCG}})$, $\dot{\rho}_{VMCG}$
will also tend to zero.\\

Then the same argument as of GCCG should be followed here too and
the classification gives:\\

(a) NEC is violated $\Rightarrow
\rho_b=\rho_{min}=\frac{3}{a_{b}^2}$,
$\rho_{\infty}=0$.\\

(b) NEC is satisfied $\Rightarrow
\rho_b=\rho_{max}=\frac{3}{a_{b}^2}$,
$\rho_{\infty}=\rho_{min}=0$.\\

where $\rho_{\infty}$ denotes the energy density evaluated in the
future. Figs. 27-29 show the dynamical behavior of the scale
factor $a$, the sum of the energy density and pressure
$(\rho_{_{VMCG}}+p_{_{VMCG}})$ and the energy density
$\rho_{_{VMCG}}$ during the bounce period for $k=1$, $\alpha=0.5$,
and $a_b=1$ for different values of $n$.\\

\subsection{\normalsize\bf{NEC in presence of a scalar field}}

To study the NEC in both the models, we write the sum of the total
energy density and pressure as

\begin{equation}\label{total energy in scalar fld}
\rho_T+p_T=(\rho+p)+(\rho_{\phi}+p_{\phi})
\end{equation}

where

\begin{equation}
\rho_{\phi}=\frac{1}{2}{\dot{\phi}}^2+V(\phi)~~~~p_{\phi}=\frac{1}{2}{\dot{\phi}}^2-V(\phi)
\end{equation}

From equations \eqref{fld eqn scalar fld 1} and \eqref{fld eqn
scalar fld 2}, we obtain at the time of bounce

\begin{equation}
(\rho_T+p_T)_{t_b}=2\left[\frac{k}{a_{b}^2}-\frac{\ddot{a_b}}{a_b}\right]
\end{equation}

From this equation, NEC is violated automatically for $k=0$ and
$k=-1$. For $k=1$, NEC is violated if $\ddot{a_b}>\frac{1}{a_b}$.
In this case, let us substitute
$\rho_{_{GCCG}}+p_{_{GCCG}}=\frac{(y^{1+\alpha}-C)-(y^{1+\alpha}-C)^{-\omega}}{y^{\alpha}}$,
$\rho_{_{VMCG}}+p_{_{VMCG}}=\frac{(1+A)y^{1+\alpha}-B_0}{{a^n}y^{\alpha}}$
and $\rho_{\phi}+p_{\phi}={\dot{\phi}}^2$ into equation
\eqref{total energy in scalar fld} to get the expressions of
$(\rho_T+p_T)_{t_b}$ for the GCCG and VMCG models as

\begin{equation}
(\rho_T+p_T)_{t_b}=\frac{({y_b}^{1+\alpha}-C)-({y_b}^{1+\alpha}-C)^{-\omega}}{{y_b}^{\alpha}}+{g_b}^2
\end{equation}

and

\begin{equation}
(\rho_T+p_T)_{t_b}=\frac{(1+A){y_b}^{1+\alpha}-B_0}{{{a_b}^n}{y_b}^{\alpha}}+{g_b}^2
\end{equation}

Therefore, for the violation of total NEC for both the models at
the time of bounce we need

\begin{equation}
C+({y_b}^{1+\alpha}-C)^{-\omega}>{y_b}^{\alpha}[y_b+{g_b}^2]
\end{equation}

and

\begin{equation}
B_0>{y_b}^{\alpha}[(1+A)y_b+{a_b}^n {g_b}^2]
\end{equation}

\section{\normalsize\bf{Discussions}}

In this work, we have considered the non-flat FRW model of the
universe where bounce occurs and the universe is filled with GCCG
or VMCG. At first, we studied the stability analysis through
dynamical system for both models and found the critical points in
flat, open and closed universe. For $k=-1,~~\alpha=1/2$, in GCCG
model, we have not got any stable critical points assuming
$\omega=-0.5$ and $C=1$. Similar solution appeared in VMCG model
as well when we chose $A=0, n=2$ and $B_0=0.5$. For GCCG and VMCG
model, the graphs of $a$ vs $H$ are shown in Figs. 1 and 2, which
are clearly not stable centers. Also in this case, $a$, $H$,
$\rho$ vs $t$ are shown in Figs. 3-8 respectively to show the
behaviour of the system. For $k=0,~~\alpha=1/2$, in GCCG model, we
got stable centers assuming $\omega=-0.5$ and $C=1$. Similar
solution appeared in VMCG model as well when we chose $A=0, n=2$
and $B_0=0.5$. These two stable solutions for the two models are
plotted in Figs. 9 and 10. For $k=1$, in GCCG and VMCG models, the
graphs of $a$ vs $H$ are shown in Figs. 11 and 12, which again are
unstable in nature.\\

The dynamical behaviour of the scale factor and the Hubble
parameter in both models have also been analysed after introducing
a scalar field. Initially when $g(t)=0$, the scale factor is seen
to have a single bounce without any oscillation for both GCCG and
VMCG models. Figs. 13 and 14 show this behavior for the two
models. Later in presence of the scalar field, i.e, when
$g(t)=e^{-\lambda t}$, the dynamical behavior of the scale factor
has shown neither a single bounce nor any oscillating behaviour.
This has been clearly shown in Figs. 16-23 with different values of $k$.\\

Finally, the energy conditions for both the models in bouncing
universe have been investigated. The criteria for bounce with or
without the violation of NEC were mentioned analytically and shown
graphically for the two models. For GCCG model, Figs. 24-26 show
the dynamical behavior of the scale factor $a$, the sum of the
energy density and pressure $(\rho_{_{GCCG}}+p_{_{GCCG}})$ and the
energy density $\rho_{_{GCCG}}$ during the bounce period for
$k=1$, $\alpha=0.5$, $\omega=-0.5$ and $a_b=1$ for different
values of $C$ and $D$. It can be clearly seen that for different
choices of the parameters $C$ and $D$, a single bounce always
appear and that never requires the violation of NEC. For VMVG
model, Figs. 27-29 show the equivalent plots showing the dynamical
behavior of the scale factor $a$, the sum of the energy density
and pressure $(\rho_{_{VMCG}}+p_{_{VMCG}})$ and the energy density
$\rho_{_{VMCG}}$ during the bounce period for $k=1$, $\alpha=0.5$,
and $a_b=1$ for different values of $n$. In this case, however a
single bounce happens to exist with late time
violation of NEC.\\\\

{\bf Acknowledgement}: The authors are thankful to IUCAA, Pune,
for their warm hospitality and excellent research facility where
most of the work has been done during a visit under the Associateship
Programme. \\\\

\end{document}